\documentclass{emulateapj}
\usepackage{amsmath}

\usepackage{ulem}
\usepackage{cancel}

\def\apj{\textit{Astrophys. J.}}                         
\def\aj{\textit{Astron. J.}}                   
\def\apjl{\textit{Astrophys. J., Lett.}}        
\def\apjs{\textit{Astrophys. J., Suppl. Ser.}}  
\def\aap{\textit{Astron. Astrophys.}}          
\def\mnras{\textit{Mon. Not. R. Astron. Soc.}} 
\def\nat{\textit{Nature}}              

\renewcommand{\S}{Section\phantom{ }}

\shorttitle{The Birth of the Early-Type Galaxies}
\shortauthors{Feldmann, Carollo \& Mayer}

\journalinfo{published in The Astrophysical Journal, 2011}
\submitted{Received 2010 December 18; accepted 2011 May 9}

\begin{document}

\title{The Hubble Sequence in Groups: \\ The Birth of the Early-Type Galaxies}
\author{R. Feldmann\altaffilmark{1,2,3}, C. M. Carollo\altaffilmark{1}, and L. Mayer\altaffilmark{1,4}}  
\altaffiltext{1}{Department of Physics, ETH Zurich,  8093 Zurich, Switzerland}
\altaffiltext{2}{Center for Particle Astrophysics, Fermi National Accelerator Laboratory, Batavia, IL 60510, USA; feldmann@fnal.gov}
\altaffiltext{3}{Kavli Institute for Cosmological Physics, The University of Chicago, Chicago, IL 60637 USA} 
\altaffiltext{4}{Institute of Theoretical Physics, University of Zurich, 8057 Zurich, Switzerland}

\begin{abstract}
The physical mechanisms and timescales that determine the morphological signatures and the quenching of star formation of typical ($\sim{}L*$) elliptical galaxies are not well understood.
To address this issue, we have simulated the formation of a group of galaxies with sufficient resolution to track the evolution of gas and stars inside about a dozen galaxy group members over cosmic history. Galaxy groups, which harbor many elliptical galaxies in the universe, are a particularly promising environment to investigate morphological transformation and star formation quenching, due to their high galaxy density, their relatively low velocity dispersion, and the presence of a hot intragroup medium. Our simulation reproduces galaxies with different Hubble morphologies and, consequently, enables us to study when and where the morphological transformation of galaxies takes place. The simulation does not include feedback from active galactic nuclei showing that it is not an essential ingredient for producing quiescent, red elliptical galaxies in galaxy groups. 
Ellipticals form, as suspected, through galaxy mergers. In contrast with what has often been speculated, however, these mergers occur at $z>1$, before the merging progenitors enter the virial radius of the group and before the group is fully assembled. 
The simulation also shows that quenching of star formation in the still star-forming elliptical galaxies lags behind their morphological transformation, but, once started, is taking less than a billion years to complete. As long envisaged the star formation quenching happens as the galaxies approach and enter the finally assembled group, due to quenching of gas accretion and (to a lesser degree) stripping. A similar sort is followed by unmerged, disk galaxies, which, as they join the group, are turned into the red-and-dead disks that abound in these environments.
\end{abstract}

\keywords{galaxies: elliptical and lenticular, cD -- galaxies: evolution -- galaxies: groups: general -- galaxies: interactions}

\section{Introduction}

Galaxy groups in the local universe harbor a large population of massive galaxies with elliptical morphology and quenched star formation \citep{2009arXiv0909.2032K}.
Observations indicate that, to a large extent,  this results from environmental forcing over the past eight billion years since redshift $z\sim~1$ \citep{2010arXiv1003.4747P}. Many physical processes may be responsible for making elliptical morphologies and quenching star formation; major and minor merging, ram pressure stripping, accretion shocks, and removal of hot gas are among those that have been suggested as possible mechanisms \citep{1972ApJ...176....1G, 1980ApJ...237..692L, 2000Sci...288.1617Q, 2000MNRAS.318..703B, 2006ApJ...650..791C, 2003MNRAS.345..349B, 2008ApJ...672L.103K, 2008MNRAS.383..593M, 2009MNRAS.397..802H}. The challenge is to connect the large-scale, cosmological formation of the group, which affects the global evolution 
of group members, to the galactic sub-kpc scale, where star formation and stellar feedback dominate the physics of the gas and the galaxy morphology \citep{2005MNRAS.361..776S, 2007MNRAS.374.1479G}. 

Both numerical simulations of the galaxy population in a cosmological context(e.g., \citealt{2008ApJ...672L.103K,2009MNRAS.399..650S, 2009MNRAS.400...43C}) and semi-analytical models of galaxy formation (e.g., \citealt{2010MNRAS.402.2321B,2010MNRAS.401.1099H, 2010MNRAS.402.1693H}) have, however, lacked so far the ability to describe at once the various environmental processes that can affect the baryonic component of galaxies, due to lack of resolution or the incomplete knowledge of the interplay between such processes. Therefore,  despite the observational evidence  that galaxy evolution  is affected by the galactic environment, how this happens  within the cold dark matter  scenario for the formation of cosmic structure is highly debated today. This notwithstanding the results of detailed, small scale numerical models, abstracted from the large-scale cosmological context (e.g., \citealt{2000Sci...288.1617Q, 2006ApJ...650..791C}), have repeatedly shown how mergers, tidal interactions, ram pressure stripping of the cold and hot gas phase as well as gas accretion from the surrounding intergalactic medium can change the morphology of individual galaxies, and alter their star formation rates (SFRs) and colors.

Recent progress in the modeling of the formation of galaxies in the cosmological context 
has finally shown that realistic galaxies can be formed in the cold dark matter model once galaxies are resolved with $\sim{}10^5$ baryonic particles and at a sub-kpc resolution \citep{
2007MNRAS.374.1479G, 2007ApJ...658..710N, 2011arXiv1103.6030G}, but this has only been achieved so far for individual galaxies in low-density (field) environments.

Therefore, key questions  remain when, and how, the disks are transformed into spheroids, star formation is quenched, and stellar mass is accreted in the massive galaxies that end up residing preferentially in the dense environments at $z=0$. We address these key questions with the help of a cosmological hydrodynamical simulation that follows the concurrent evolution of many galaxies before, during, and after they join a relatively dense,  group-sized virialized region with virial radius $R_\mathrm{200}=396$ kpc and virial mass $M_\mathrm{200}=1.1\times{}10^{13}$ $M_\odot$ at $z\sim{}0$.

The structure of this paper is as follows. We describe our simulation and analysis methods in \S \ref{sect:Sim} and \ref{sect:Analysis}, respectively. In \S \ref{sect:Results} we present and discuss our main results, including the properties of the group satellite galaxies (\S \ref{sect:Hubble} and \ref{sect:scaling}), and the origin of the morphological and photometric transformations (\S \ref{sect:Drivers}). We conclude in \S \ref{sect:Conclusions}.

\section{Simulation details}
\label{sect:Sim}


The simulation set-up and the employed zoom-in technique is described in detail elsewhere \citep{2010ApJ...709..218F}. In brief, a group halo is selected from a dark matter simulation of a 123 Mpc periodic box. Higher resolution initial conditions centered on the group with WMAP-3 cosmological parameters \citep{2007ApJS..170..377S} are generated using \texttt{grafic-2} \citep{2001ApJS..137....1B}. The simulation is run with the parallel TreeSPH code \texttt{GASOLINE} \citep{2004NewA....9..137W}. The mass (force) resolution of the simulation is $6.4\times{}10^6$ $M_\odot$ (0.5 kpc), $1.4\times{}10^{6}$ $M_\odot$ (0.3 kpc), and $4\times{}10^{5}$ $M_\odot$ (0.3 kpc) for dark matter, gas and star particles, respectively, which allows us to resolve the half-mass radii of the massive group members  ($\sim{}1$ kpc). We output 92 snapshots between $z=41.5$ and $z=0.1$ equidistant in time, i.e., with $\Delta{}t\approx{}130$ Myr.

Besides following gravitational, hydrodynamical, and optically thin radiative processes of matter in an expanding universe directly, the simulation includes modeling of star formation, feedback from type Ia and type II supernovae, mass loss by stellar winds, and metal enrichment. The simulation parameters were originally tuned to produce disk galaxies in galactic halos \citep{2006MNRAS.373.1074S, 2007MNRAS.374.1479G}. We use a star formation efficiency per free fall time of 0.05, and allow stars to form in a probabilistic fashion if the density exceeds 0.1 cm$^{-3}$, the gas temperature is lower than 15,000 K, the gas is in an overdense ($\delta{}>55$) region, and part of a convergent flow. A thermal feedback of $4\times{}10^{50}$ erg is injected into the gas per supernova. The cooling of the gas is turned off for the time corresponding to the end of the snowplow phase of a supernova type II blast wave \citep{2006MNRAS.373.1074S}. The cooling is not disabled for supernovae of type Ia. The simulation includes stellar mass loss due to stellar winds and a spatially uniform UV background \citep{1996ApJ...461...20H}. The simulation we present here corresponds to the run G2-HR in \cite{2010ApJ...709..218F} where it has been used to analyze the properties of a massive, central group galaxy.

As we show in \S \ref{sect:scaling}, the same physical prescriptions produce, when used to simulate a relatively massive potential of a group of galaxies, $z\sim{}0$ galaxies with a broad range of luminosities, sizes and kinematics, which are roughly consistent with the properties of disk and elliptical galaxies in the local universe. They are also adequate to produce massive, early-type galaxies at the centers of galaxy groups \citep{2010ApJ...709..218F}.

\section{Halo extraction and the measurement of their properties}
\label{sect:Analysis}

Halos and halo centers are identified with the AMIGA HaloFinder (AHF; \citealt{2004MNRAS.351..399G, 2009ApJS..182..608K}), that is capable of detecting sub-halos (our satellites) within halos (our group). Halos extend out to the virial radius $R_\mathrm{200}$, which we define as the radius enclosing a mean matter density of $200/\Omega_m$ times the mean density of the universe, or out to the tidal radius, whichever is smaller. The virial mass $M_\mathrm{200}$ is the mass within $R_\mathrm{200}$. We then use the following technique to track satellites over time. First, we identify each halo $h$ in each snapshot $s$ below $z=4$ that $(1)$ has a bound mass of more than $1.4\times{}10^{10}$ $M_\odot$, $(2)$ contains more than 50000 particles, and $(3)$ lies within 630 comoving kpc from the group center. Second, for each such determined halo $h_s$ we determine its main progenitor $h_{s-1}$ in the previous snapshot and its main successor $h_{s+1}$ in the following snapshot and construct a complete timeline $(h_0,\ldots,h_s,\ldots,h_n)$ by recursion. Let $N_1$, $N_2$ and $N_\mathrm{common}$ be the number of particles in halo $h_1$, the number of particles in halo $h_2$ and the number of particles common to both halos, respectively. We then use the ratio $\eta(h_1,h_2)=N^2_\mathrm{common}/(N_1N_2)$ to define main successors and progenitors. Specifically, we define the halo with the largest $\eta$ from snapshot $s+1$ (any halo is in principle eligible, not only the ones selected as described above) as the main successor of halo $h_s$ and the halo with the largest $\eta$ from snapshot $s-1$ as the main progenitor (but
excluding the parent group halo). Since the group halo $h_G$ is a potential main successor this algorithm sometimes finds multiple timelines for a given halo $h_s$. For instance, the following timelines of halo $h_s$ could be produced: $(h_0,\ldots,h_s,h_{s+1},\ldots,h_n)$, $(h_0,\ldots,h_s,h_{s+1},h_G,\ldots,h_G)$, and $(h_0,\ldots,h_s,h_G\ldots,h_G)$. In this case, we only keep the timeline that traces the halo $h_s$ as long as possible as an entity separate from the group halo (which is the first timeline in the example). Following this procedure, we obtain $\sim{}20$ satellites of which 13 have not merged with the central galaxy by $z=0.1$.

Unless noted otherwise SFRs are measured within a sphere of 20 physical  kpc. The specific SFR is obtained by dividing the SFR by the stellar mass. Stellar masses are defined as the mass of the stellar component that is bound to the halo (using AHF) within a sphere of 20 physical kpc. 
This radius encloses almost all ($\sim{}98$\% on average, 91\% for the largest galaxy at $z=0.1$; $\sim{}100\%$ at $z\gtrsim{}1$) of the stellar mass of non-central group members within their virial (or truncation) radius.

Kinematic properties are determined from a slit that is 24 kpc long, 2 kpc wide and contains 24 bins along the two-dimensional major axis of an edge-on view of the galaxy. The rotation velocities and dispersions are determined from the moments of the line-of-sight velocities. 
In Figures \ref{fig:Sat} and \ref{fig:Vrot} we measure the moments of the line-of-sight velocities on inclined galaxies with an inclination\footnote{\cite{2007AJ....134..945P} measure the SDSS $i$-band Tully--Fisher relation, see Figure \ref{fig:Sat}, on a sample of disk galaxies with a median inclination of $i=72^{\circ}$.} of $i=70^{\circ}$ and then correct for the inclination by diving the velocities by $\sin(i)$.
Maximal rotation velocities $v_\mathrm{rot}$ are taken to be half the difference between the maximum and the minimum rotation velocity along the slit. 
Central velocity dispersions $\sigma_\mathrm{cen}$ are measured within a 1 kpc aperture. The measured value is increased by a factor 1.1 to account for the radial decline of the velocity dispersion \citep{1993MNRAS.265..553C, 2001MNRAS.326..221T}. The quantity $v/\sigma$ refers to the ratio of maximum rotation velocity and central velocity dispersion.

Cold gas is defined as the gas with a temperature of less then $3.2\times{}10^{4}$ K, unless noted otherwise. The neutral hydrogen mass is derived from the hydrogen mass fraction that is calculated within GASOLINE during the simulation. Our simulation does not include an explicit treatment of molecular hydrogen. Hence, in the analysis we do not distinguish between atomic and molecular hydrogen components.
Colors (all in the AB-system) are derived from stellar synthesis models, typically from the models of \cite{2003MNRAS.344.1000B}, but we have also tested the single stellar populations of \cite{2005MNRAS.362..799M} finding that the latter reddens the $B-I$ color at $z=0.1$ of galaxies with $B-I\lesssim{}1.1$ by up to 0.1 mag. Hence uncertainties in the stellar population models are unlikely to have strong effects on our results, but can nonetheless change colors and magnitudes by a few percent.

We reduce the effect of star formation in the central region on the colors by fixing the mass-to-light ratio from within one softening length to its value at one softening length. In order to estimate the effect of a very young ($<100$ Myr) stellar component, we have tested a simplified version of the dust-obscuration model of \cite{2000ApJ...539..718C}, namely, reducing the flux of the very young stellar particles to zero as done in, e.g., \cite{2007ApJ...658..710N}, but not applying any further dust corrections. We find that galaxies with $B-I$ colors below $\sim{}1$ are typically reddened by $0.05$-$0.1$ mag and redder galaxies are affected on a less than 0.05 mag level. Given that the corrections are overall rather small we will refer to the single stellar populations of \cite{2003MNRAS.344.1000B} without a flux reduction of very young stars.

The smoothed particle hydrodynamics technique allows us to follow the temperature history of individual gas parcels. We track the temperature history of each gas particle over all snapshots in order to assess whether accreted gas has been cooled from a hot halo or whether it stayed cold since the beginning of the simulation. We 
determine for each gas particle its maximum temperature over its history, but ignore the times when a gas particle is  less than 30 kpc away from the satellite center in order to avoid that its maximum temperature is set by supernova feedback, see \cite{2009ApJ...694..396B}. We identify gas particles with a maximum temperature above (below) $2.5\times{}10^5$ K as contributors to hot (cold) accretion \citep{2005MNRAS.363....2K, 2009MNRAS.395..160K}. Despite its simplicity this temperature-based split into hot/cold accretion gives similar results compared to more sophisticated treatments \citep{2009ApJ...694..396B}. Since each star particle is spawned from a unique gas particle we can also use the maximum temperature of the spawning gas particle to asses whether a star particle has been formed from gas that was accreted hot or cold.

To obtain the net, instantaneous accretion rates of gas and stars onto a satellite galaxy we first determine its center-of-mass velocity from all particles within a sphere of 20 kpc. Radial velocities $v^r$ of gas and star particles are then computed in the rest frame of the satellite. We then estimate the net mass flux of, e.g., gas with a maximum temperature below $2.5\times{}10^5$ K, by summing over all particles of the relevant type in a spherical shell at $r=35$ kpc distance from the center
\[
dM/dt\approx{}4\pi{}r^2\rho{}\sum_im_iv^r_i/\sum_im_i=4\pi{}r^2\sum_im_iv^r_i/V.
\]
Here, $\rho$ and $V$ are the mean density and volume of the spherical shell. Its thickness is 10 kpc to ensure that it contains at least several hundred particles. For the accretion of bound material we only include the particles identified as bound to the satellite by AHF, otherwise we use all the particles in the shell.

We quantify the mass loss of cold gas due to consumption and stripping by tracing forward in time the cold gas particles that are contained within 20 kpc at a snapshot $\sim{}100$-$200$ Myr before infall time. We consider that an initially cold gas particle may be kept as gas particle within a 20 kpc distance (``kept''), that it may be kept after being transformed into a star particle (``consumed''), or that it leaves the 20 kpc radius (``stripped''). A small, but non-negligible fraction of the ``kept'' gas is heated, e.g., by supernova feedback. However, to first order the heating and cooling of gas should balance and hence the heated gas mass will be replaced by a similar gas mass of cooled gas. Hence, the ``kept'' gas mass should roughly correspond to the kept cold gas mass. The stated mass fractions and errors refer to the mean and the standard deviation of the various mass fractions for all $z=0.1$ gas-poor satellites.

\begin{figure}
\begin{center}
\begin{tabular}{c}
\includegraphics[width=85mm]{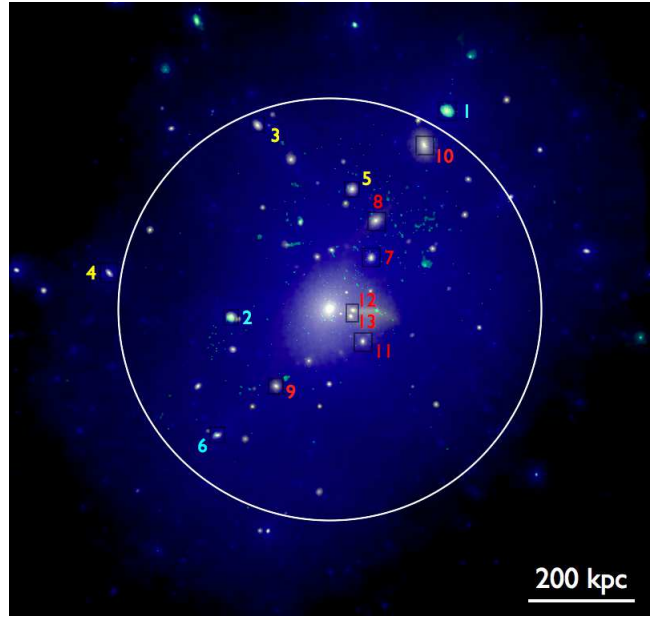}
\end{tabular}
\caption{Simulated group at $z=0.1$. A massive central galaxy is surrounded by a large number of massive satellite galaxies. The image is 1.2 Mpc $\times{}$ 1.2 Mpc across and shows the projected mass densities of dark matter (blue; from 1.45 to 1450 $M_\odot$ pc$^{-2}$), neutral hydrogen (green) and stellar matter (yellow; both from 0.36 to 365 $M_\odot{}$ pc$^{-2}$). The white circle indicates the virial radius $R_\mathrm{200}=396$ kpc of the group. The properties of the satellites with a stellar mass $>10^{10}$ $M_\odot$ (numbered) are shown in Figure~\ref{fig:GroupCensus}. The galaxies are either entering the group for the first time (e.g., galaxy 1), or orbit within the virial radius of the group, or already crossed the group and are now at an apocenter (e.g., galaxy 4).\label{fig:GroupOverview}}
\end{center}
\end{figure}

\begin{figure*}
\begin{center}
\includegraphics[width=145mm]{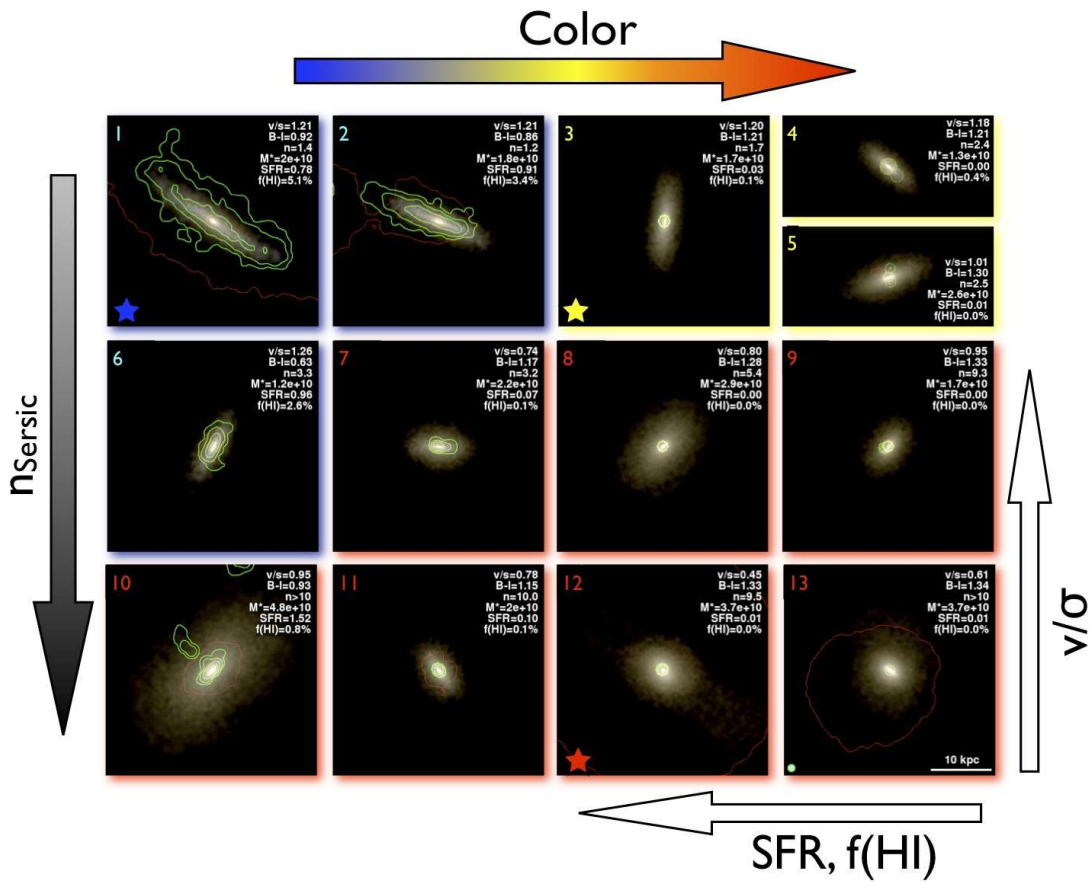}
\caption{{\footnotesize Massive ($M>10^{10} M_\odot$) galaxies orbiting within the group at $z=0.1$. Each composite image shows an edge-on view,  i.e. a view along the intermediate axis of the reduced moment-of-inertia tensor of the stellar component within 10 kpc, of a satellite galaxy. The galaxies are arranged with respect to their color, Sersic index, rotational support, star formation rate, and gas fraction. Fundamental galaxy properties are shown at the top right of each panel ($v/\sigma$, $B-I$ color, Sersic index, stellar mass in $M_\odot$, SFR in $M_\odot$ yr$^{-1}$, HI fraction). Galaxies 1, 2 and 6 are classified as normal disk galaxies, $3-5$ as gas-poor disks and $7-13$ as elliptical galaxies. A representative of each class is marked with a colored star. (Red star) This galaxy has a  red color, hosts virtually no neutral gas, is almost spherical (sphericity of the stellar component $c/a=0.77$), has a steep surface mass profile, and its  kinematics is dominated by velocity dispersion. (Yellow star) The galaxy is a red, gas-poor, rotating disk galaxy ($c/a=0.38$) with an $I$-band exponential scale length $a_I=1.4$ kpc. (Blue star) This galaxy hosts $\sim{}10^9$ $M_\odot$ of HI, has a bluer $B-I$ color, and harbors a rotating stellar disk ($c/a=0.36$) with $a_I=2.3$ kpc. The RGB color channels of the images correspond to the surface brightness in the rest-frame Bessel $B$, $R$ and $I$ filter bands \citep{1990PASP..102.1181B}, respectively, and range from 16 mag arcsec$^{-2}$ to 24.5 mag arcsec$^{-2}$. Green contour lines indicate the column densities of neutral hydrogen corresponding to 0.1, 1, 10, and 100 $M_\odot$ pc$^{-2}$, while thin (thick) red contours correspond to column densities of  1 (10) $M_\odot$ pc$^{-2}$ of bound hot ($T>2.5\times{}10^5$ K) gas. The asymmetric gas contours in the two top left panels reveal the action of ram-pressure as the satellites move through the intragroup medium. The scale of all images is indicated in the bottom right panel. The green circle at the bottom-left corner of panel 13 has a radius of two gravitational softening lengths and indicates the resolution limit of the simulation.}
\label{fig:GroupCensus}}
\end{center}
\end{figure*}

We measure single component Sersic indices $n_\mathrm{sersic}$ on face-on and edge-on mock images with GALFIT \citep{2002AJ....124..266P}. In particular, we compute noise (``sigma'') maps based on the particle-per-pixel number in mock images of the surface mass density assuming a Poisson statistics. We exclude the central three softening lengths from the fit and use the average value of the face-on and edge-on $n_\mathrm{sersic}$ value. 
This approach substantially reduces the spurious effects of  ``overcooling'' (e.g., \citealt{2008ASL.....1....7M}) in the morphological  classification, as it puts weight  on the mass profile outside the very central regions. Our classification criterion, albeit coarse, allows us nonetheless to distinguish between the disk and elliptical satellites which are discussed in the main text.

\section{Results}
\label{sect:Results}

\subsection{A Hubble Sequence in a Galaxy Group}
\label{sect:Hubble}

\begin{figure*}
\begin{center}
\begin{tabular}{cc}
\includegraphics[width=80mm]{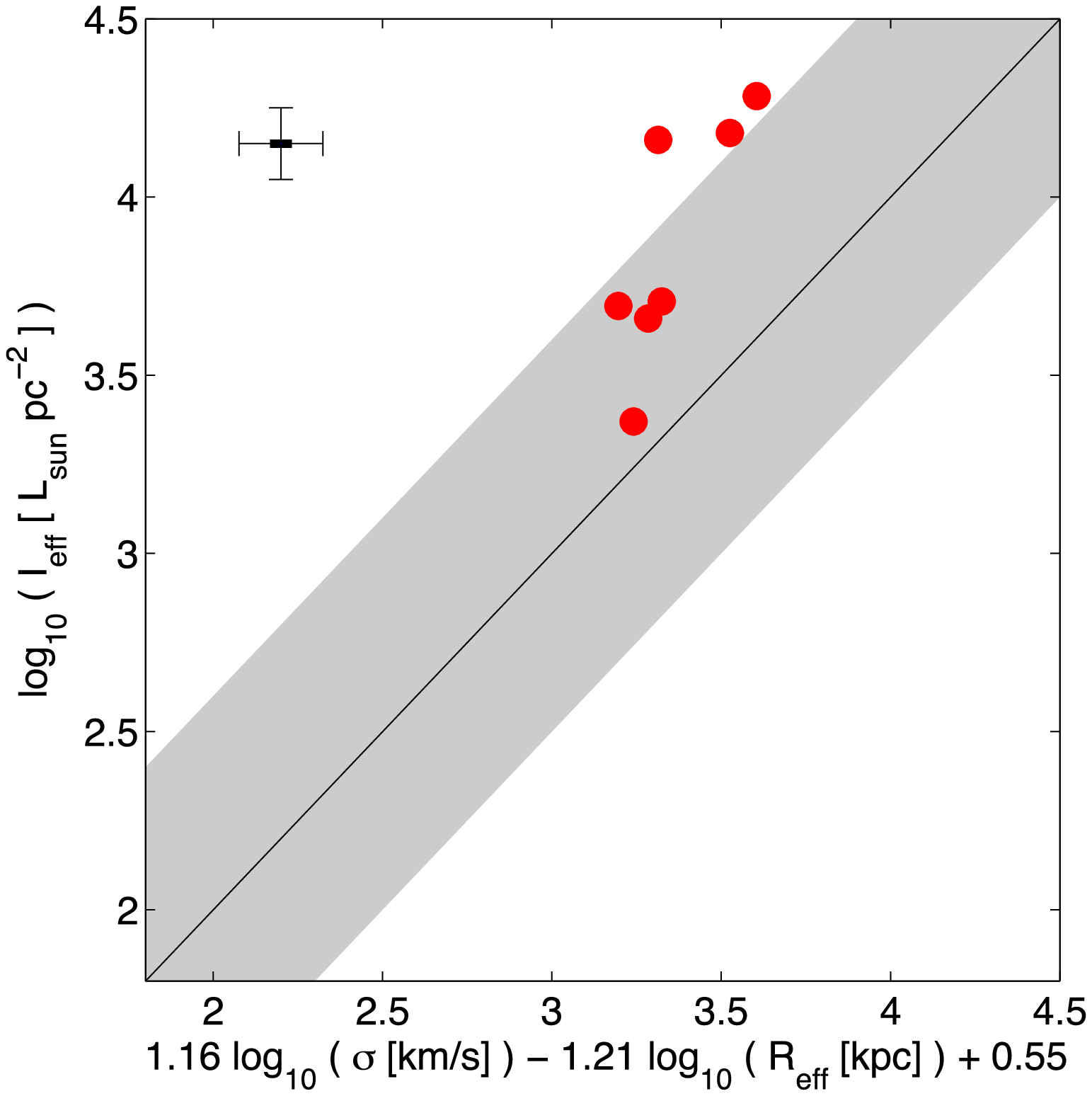}
\includegraphics[width=75mm]{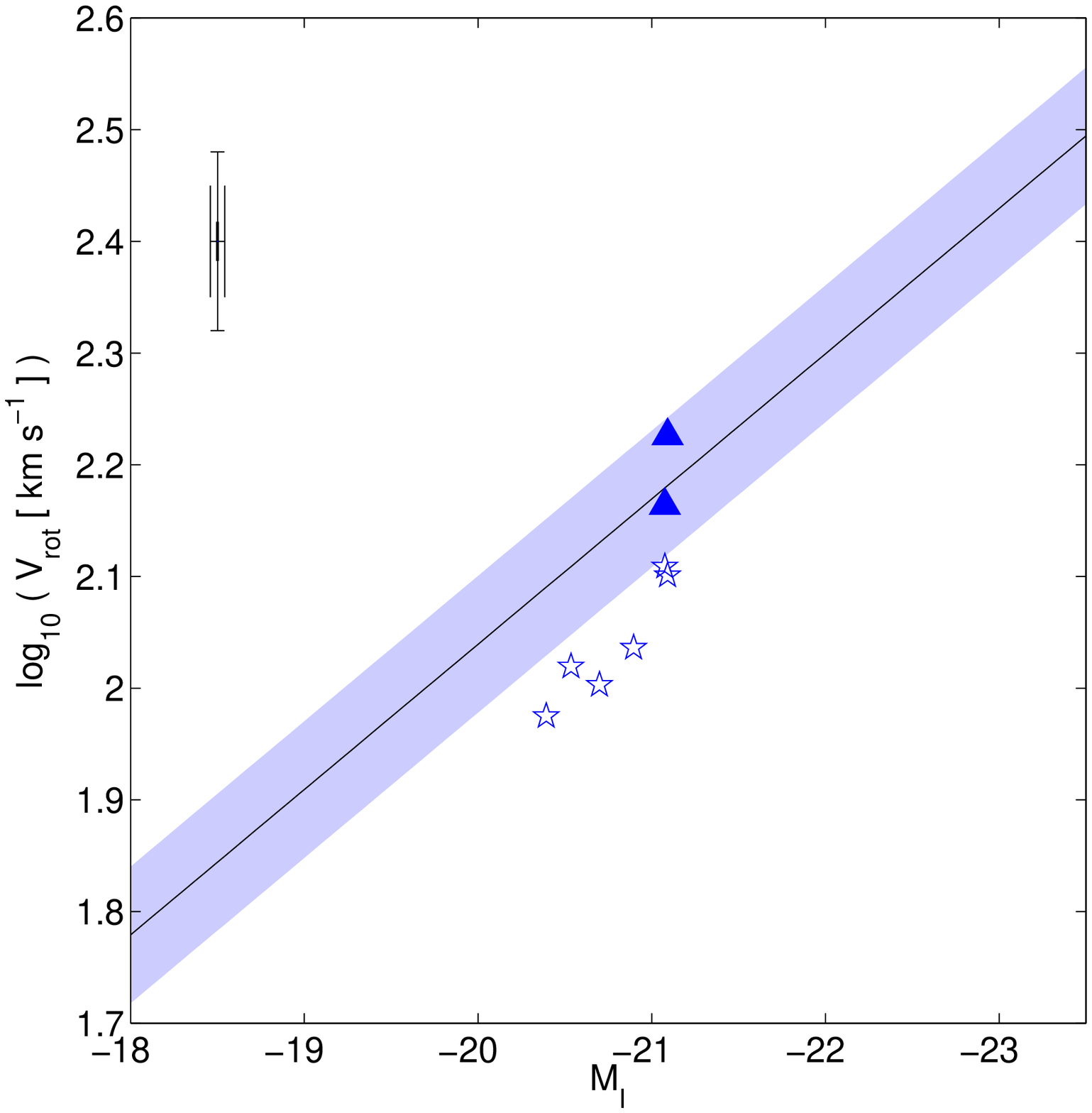}\\
\end{tabular}
\caption{Comparison between simulated and observed galaxy scaling relations. Simulated values are inferred from the $z=0.1$ snapshot; blue (empty) and red (filled) symbols  correspond to satellites that are classified as  disks and ellipticals, respectively. Left: projection of the fundamental plane of early-type galaxies \citep{2009ApJ...698.1590G}. The gray stripe indicates the location of real galaxies and red dots are our simulated objects with an early-type morphology by redshift $z=0.1$.
Right: the Tully--Fisher (TF) relation. Star symbols show maximum \emph{stellar} line-of-sight rotation velocities of the disk satellites; triangles show the maximum line-of-sight rotation velocities of neutral hydrogen for the two disk satellites with $f_\mathrm{gas}>3\%$. The solid line shows the $i$-band TF relation from the SDSS survey \citep{2007AJ....134..945P} and the shaded area indicates its scatter. HI rotation velocities and $i$-band luminosities of the satellites are consistent with the TF relation. We verified (not shown) that this result also holds for dust-corrected $B$-band luminosities \citep{2001A&A...370..765V, 2001ApJ...563..694V, 2001ApJ...550..212B}. Error bars in the top left corners of each figure include  statistical errors for individual satellites (including Poisson noise in the galaxy representations, binning errors and fitting errors), and an estimate of the systematic errors in converting simulated data into observables. These systematic errors include contributions from the fixed aperture measurements used to measure the half-light radii, from estimating the luminosities from stellar synthesis models,  from neglecting dust  effects, as well as from differences between kinematic data extracted from moments and from  Gauss--Hermitian fits.
\label{fig:Sat}}
\end{center}
\end{figure*}

The simulation was evolved down to  redshift  $z=0.1$. At this epoch, 13 well-resolved galaxies, with stellar masses $>10^{10}$ $M_\odot$, orbit in, or near, the galaxy group, see Figure \ref{fig:GroupOverview}. The stellar mass budget of the group is dominated by an early-type galaxy at its center, whose properties are discussed in detail elsewhere
 \citep{2010ApJ...709..218F}. The mass range of these satellite galaxies straddles across the transition mass of $\sim{}3\times{}10^{10}$ $M_\odot$ below (above) which disk (early-type) galaxies are more numerous than galaxies with early-type (disk) morphology.
Indeed, Figure \ref{fig:GroupCensus} shows that, at $z=0.1$,  the simulated satellite galaxies span a broad range in morphological properties, from very flat disks  to almost spherical spheroids, as well as in rotational support ($v_\mathrm{rot}/\sigma_\mathrm{cen}\sim{}0.4-1.3$), color ($B-I=0.6-1.3$ mag), neutral hydrogen fraction ($f_\mathrm{gas}\sim0\%-5\%$), and SFRs. Properties range from passively evolving galaxies to moderately star-forming systems, with SFRs that are  typical of similar mass $z\sim{}0$ disks, i.e., $\sim{}1$ $M_\odot$yr$^{-1}$. 

Bulge to disk decompositions remain challenging even at the sub-kpc resolution of our simulation. However, based on standard morphological classification criteria that use global galactic properties such as the ratio of stream-to-dispersion velocities, and the index of single-component Sersic fits to the surface brightness profiles, we identify six group members as rotationally supported, disk galaxies ($v_\mathrm{rot}/\sigma_\mathrm{cen}\geq{}1$, $n_\mathrm{Sersic}<2.5$). These galaxies show a range of colors and SFRs that range from those of normal star-forming spirals  to those of passive spirals/S0 galaxies.  
Specifically, three of the disk galaxies have red colors ($B-I\sim{}1.2-1.3$), contain no significant amount of neutral hydrogen ($<1\%$), and have stopped forming stars (top right in Figure \ref{fig:GroupCensus}). The three other disk galaxies, instead, have bluer colors ($B-I\sim{}0.6-0.9$), significant neutral hydrogen fractions ($\gtrsim{}3\%$), and SFRs of $\sim{}1$ $M_\odot$ yr$^{-1}$ (top left in Figure \ref{fig:GroupCensus}). The remaining seven satellites have properties that are typical of early-type (hereafter ``elliptical'') galaxies, i.e., $n_\mathrm{Sersic}\geq{}2.5$ and  $v_\mathrm{rot}/\sigma_\mathrm{cen}<1$.

\subsection{Scaling Relations of the Simulated Satellite Galaxies}
\label{sect:scaling}

Real galaxy populations are observed to obey  fundamental scaling laws, such as the fundamental plane (FP) relation \citep{1987ApJ...313...42D, 1992ApJ...399..462B} and the Tully--Fisher (TF) relation \citep{1977A&A....54..661T}.  The FP  for elliptical galaxies shows that most such galaxies populate a thin two-dimensional surface in the three-dimensional space of  characteristic stellar velocity $\sigma$, half-light radius $r_\mathrm{eff}$ , and surface brightness within the half-light radius, $\Sigma_\mathrm{eff}$. The FP reflects the virial relation between masses, sizes, and velocities; the observed tilt relative to the virial plane is interpreted as a dependence of the mass-to-light ratio with galaxy mass. Similarly, the TF relation connects the luminosity of disk galaxies and their rotation velocities. Observations indicate $L\propto{}V^\alpha_\mathrm{rot}$ with $\alpha=3-4$ depending on wavelength and how dust extinction is modeled.

We use these scaling laws to assess how similar, in their global properties, our simulated objects are to real galaxies. For this purpose, we extract from the last simulation snapshot ($z=0.1$) the following properties for the satellites in the Bessel $B$ band \citep{1990PASP..102.1181B} and the Sloan Digital Sky Survey (SDSS) $i$ band \citep{1996AJ....111.1748F}:
(1) circular effective radii $r_\mathrm{eff}$, that include half the total flux within projected 20 kpc. The flux from within one baryonic softening length is derived from the included mass and the mass-to-light ratio at one softening length. (2) The surface brightness which is proportional to the flux divided by $r^2_\mathrm{eff}$. (3) The flux-weighted stellar line-of sight velocity dispersion and rotation velocity. (4) The line-of-sight rotation velocity of neutral hydrogen for the two disk satellites with a sufficient cold gas content at $z=0.1$, measured as for the stellar component.

\begin{figure}
\begin{center}
\begin{tabular}{cc}
\includegraphics[width=80mm]{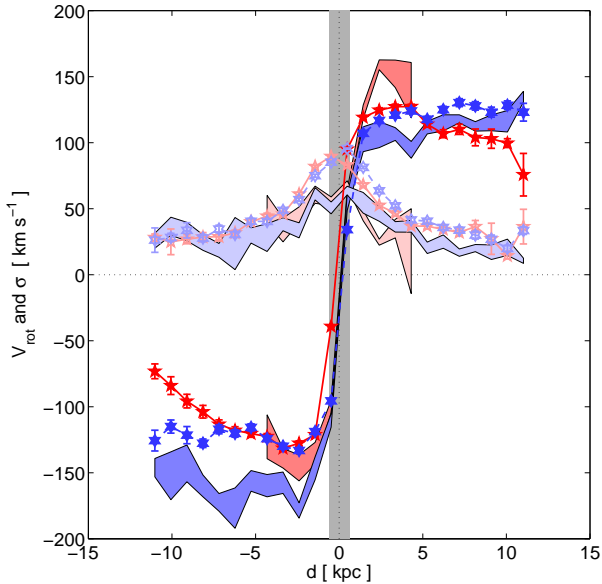}
\end{tabular}
\caption{Rotation velocities and velocity dispersions of the two $z=0.1$ disk satellites with extended gas disks as function of radius. The kinematic properties of galaxy 1(2), see Figure~\ref{fig:GroupCensus}, are shown in blue (red). Filled (empty) star symbols correspond to stellar rotation velocities (stellar velocity dispersions) along the slit. Dark (light) shaded areas show HI rotation velocities (HI velocity dispersions).  The width  of the shaded regions indicates the errors in the HI kinematic quantities. Errors are computed for each radial bin separately via bootstrapping.
\label{fig:Vrot}}
\end{center}
\end{figure}

\begin{table}[b]
\begin{center}
\caption[The Classification of $z=0.1$ Group Satellites]{The Classification of $z=0.1$ Group Satellites with Stellar Masses above $10^{10}$ $M_\odot$ and Their Main Progenitors}
\renewcommand{\arraystretch}{1.25}
\begin{tabular}{lr|rrr}
\tableline
\tableline
Satellite Classification & $z=0.1$ & $z=0.5$ & $z=1$ & $z=2$ \\ \tableline
\# Red                       & 9 & 4 & 0 & 0 \\ 
\# Blue                     & 4 & 9 & 13 & 13 \\ \tableline
\# Gas-poor             & 10 (9) & 5 (2) & 1 (1) & 0 (0) \\ 
\# Intermediate-gas     & 3 (3) & 8 (6) & 9 (3) & 2 (1)  \\ 
\# Gas-rich               & 0 (1) & 0 (5) & 4 (9) & 11 (12) \\  \tableline
\# Quiescent     & 9 & 4 & 1 & 0 \\
\# Star-forming & 4 & 7 & 6 & 1 \\
\# Strongly SF  & 0 & 2 & 6 & 12 \\  \tableline
\# Quiescent (total) & 8 & 1 & 0 & 0 \\
\# Star-forming (total)& 4 & 8 & 4 & 0 \\ 
\# Strongly SF (total) & 1 & 4  & 9 & 13 \\ \tableline
\tableline
\end{tabular}
\tablecomments{
The rest frame $B-I$ color is used to define red ($B-I\geq{}1$) and blue ($B-I<1$) galaxies.  Galaxies are defined as gas-rich if they have a 
HI-to-stellar mass ratio or (for the values given in brackets) a cold gas-to-stellar mass ratio in excess of 10\%. Gas-poor galaxies have $f_\mathrm{gas}<$1\%. Galaxies with $f_\mathrm{gas}$  in the range 1\%--10\% are referred to as intermediate-gas systems. Galaxies with specific SFRs (SSFRs) $<0.02$ Gyr$^{-1}$ are defined as quiescent galaxies, and otherwise as star forming galaxies. Galaxies are furthermore considered to be ``strongly star forming'' if their SSFR is larger than the inverse of the age of the universe, i.e., SSFR $>$ 0.08 Gyr$^{-1}$ at $z=0.1$. The classification in the last three rows differs from the one in the previous three rows due to the inclusion of the star formation occurring within the unresolved central softening length  $\epsilon_\mathrm{bar}$. Note that at $z\sim{}1-2$ most progenitors are blue, gas rich, star-forming galaxies.
\label{tab:StatGroupSat}}
\end{center}
\end{table}

\begin{figure}
\begin{center}
\begin{tabular}{cc}
\includegraphics[width=78mm]{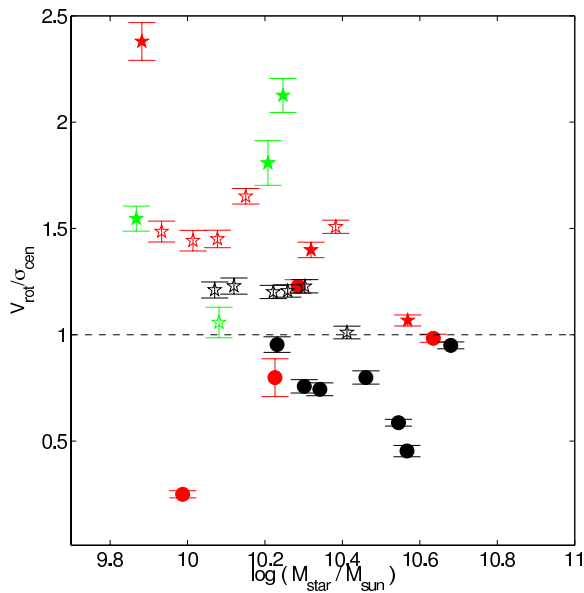}
\end{tabular}
\caption{Kinematic  properties of massive $z=0$ satellites and their progenitors. Shown is the ratio of maximum stellar line-of-sight rotation velocity and central velocity dispersion vs. stellar mass $M_\mathrm{star}(z)$. Plotted are the 13 satellites with mass in excess of $10^{10} M_\odot$ at $z=0.1$. The redshift $z$ is indicated by color: $z=0.1$ (black), $z=1$ (red), and $z=2$ (green); Circles (stars) indicate galaxies with a sphericity larger (smaller) than 0.45;  filled (empty) symbols denote galaxies that are classified at $z=0.1$ as elliptical (disk) galaxies. At $z\sim{}1-2$ most  progenitors are rotation-supported galaxies. 
\label{fig:VelSigmaMass}}
\end{center}
\end{figure}

In the left panel of Figure \ref{fig:Sat}, we analyze the position of the simulated galaxies  within the FP.  Four of the seven early-type galaxies in our simulated sample have FP properties of real galaxies of similar type. The remaining three are slightly more compact.  Numerical effects or missing physical ingredients are likely to play a role in generating this effect \citep{2003ApJ...590..619M, 2010ApJ...709..218F}.  For instance, spurious angular momentum loss due to limited resolution may increase the amount of gas that overcools in the central region \citep{2008ASL.....1....7M}. The lack of feedback from an active galactic nucleus (AGN) \citep{2010arXiv1003.4744T} and of galactic outflows \citep{2009ASPC..419..347D, 2010MNRAS.406.2325O} may increase the fraction of gas that cools and is converted into stars in the central region of the galaxy. Finally, the low star formation threshold may reduce the efficiency of the supernova feedback \citep{2010Natur.463..203G}.

In the right panel of Figure \ref{fig:Sat} we show how the rotation velocities of the simulated disk galaxies compare with the observed TF relation \citep{2007AJ....134..945P}. At $z=0.1$ only two of our six disk galaxies have extended gas disks and thus allow us to measure rotation velocities based on HI. The luminosities and rotation velocities of these two galaxies are in good agreement with the observed TF relation. The figure also demonstrates that rotation velocities derived from the stellar component are not a good proxy for the rotation velocity of the gaseous component, because a significant fraction of the kinetic energy of stars is in unordered motions. This latter point can also be seen in Figure \ref{fig:Vrot}, where we show the radial dependence of the HI and stellar kinematics, respectively, of the two disk galaxies with extended gas disks. Clearly, neutral hydrogen has a larger rotation velocity and a smaller velocity dispersion than the stellar component. The figure further shows that one of the galaxies has a ``flat'' rotation curve both in HI and in the stellar component. The gas disk of the other galaxy is highly asymmetric due to ram-pressure stripping and we can measure its HI rotation velocity out to about $\sim{}5$ kpc. At larger radii we observe a declining rotation velocity in the stellar component.

\begin{figure}
\begin{center}
\begin{tabular}{c}
\includegraphics[width=80mm]{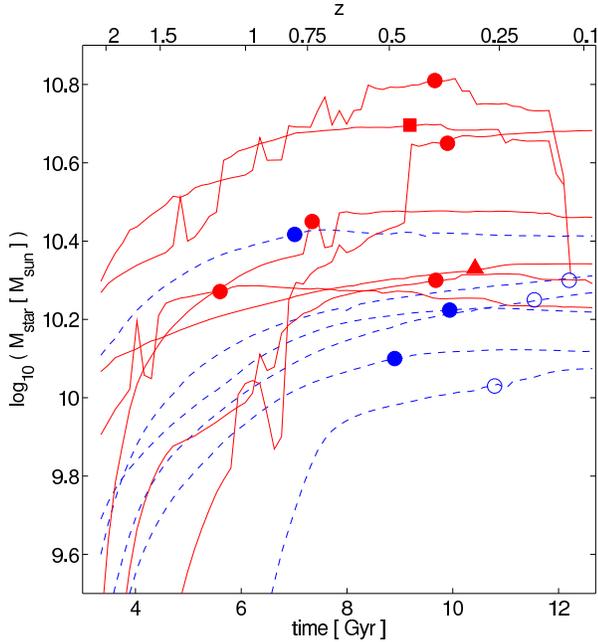}
\end{tabular}
\caption{Assembly histories since $z=2$ of all $z=0.1$ group members with a stellar mass above $10^{10}$ $M_\odot$. Red solid (blue dashed) lines show the stellar mass of $z=0.1$ elliptical (disk) satellites as function of time (bottom axis) and redshift (top axis). Filled (empty) symbols indicate the time when the progenitor of a $z=0.1$ gas-poor (intermediate-gas) satellite enters the virial radius of the group. Sudden increases in stellar mass are caused by merger events, often before infall into the group, while close pericentric passages that lead to tidal stripping reveal themselves through a step-wise stellar mass loss occurring after their entrance in  the group. The gas-poor disk satellites do not grow significant stellar mass after they enter the group, due to quenching of their star formation. One of the elliptical satellites does not show signs of significant merging since $z\sim{}2.2$ (red triangle). This galaxy assembles most of its stellar mass early on, before such epoch, i.e., much earlier than most disk satellites, and, not surprisingly, this results  in a massive, very compact spheroidal.
\label{fig:StellMass}}
\end{center}
\end{figure}

We conclude that, overall, the simulated galaxies match the fundamental scaling laws obeyed by the $\lesssim$ $L*$ $z=0$ galaxy population reasonably well. Of course, the match between real and simulated galaxies is neither expected nor observed to be yet perfect. Yet, we have come a long way in reproducing the rich diversity of the $z=0$ galaxy population, from disks to ellipticals. This is achieved with numerical recipes for star formation and stellar feedback which were  originally tuned to  produce a Milky-Way-type galaxy \citep{2006MNRAS.373.1074S, 2007MNRAS.374.1479G}.
Despite this success in reproducing galaxies with approximatively realistic properties, we should point out, however, that the stellar mass to halo mass\footnote{The latter is defined as the mass of the halo when the satellite first enters the virial radius of the group.} ratios of the group satellites are similar to the ones obtained in previous cosmological simulations of galactic halos ($\sim{}20\%-80\%$); e.g., \cite{2003ApJ...591..499A, 2007MNRAS.374.1479G, 2009MNRAS.396..696S, 2011MNRAS.410.1391A}. 
Halo abundance matching approaches based on the observed stellar mass function typically predict stellar mass to halo mass ratios that are lower by a factor of a few, e.g.,  \cite{2010MNRAS.404.1111G}. 
This mismatch may have its origin in an over-efficient conversion from gas to stars in simulations indicating that the feedback prescription is incomplete and that additional mechanisms that remove gas from the galaxies, e.g., radiation pressure \citep{2011arXiv1101.4940H}, or supernova driven outflows \citep{2011arXiv1103.6030G}, are required.

\begin{figure}
\begin{center}
\begin{tabular}{c}
\includegraphics[width=88mm]{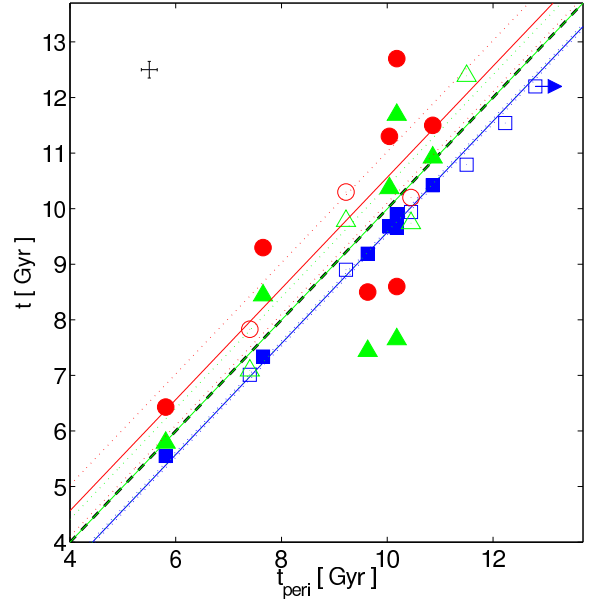} \\
\end{tabular}
\caption{Events in the life of the group satellites. The $x$-axis indicates the time of the first pericentric passage $t_{\rm peri}$. The $y$-axis indicates the following events: the time when the satellite first crosses the virial radius of the group (shown using blue squares), the time when the HI-to-stellar mass ratio drops below $1\%$ (green triangles), and the time when the $B-I$ color increases above 1 (red circles). Note that some satellites have not turned into gas-poor or red galaxies by $z=0.1$ and, therefore, for these galaxies green triangles or red circles are missing. Galaxies that are identified as disks (ellipticals) at $z=0.1$ are shown with empty (filled) symbols.
Linear fits with unity slope are shown as colored solid lines, and the fit errors as dotted lines.  The satellites approach their first pericentric passage typically $0.4\pm{}0.1$ Gyr after infall into the group and, by then, turn into gas-poor ($f_\mathrm{gas}<1\%$) galaxies. On average $0.6\pm{}0.5$ Gyr later the galaxies become red. The error of a typical data point that arises from the limited time resolution of the simulation is indicated at the top left. The arrow at the top right highlights  a gas-rich disk that has not yet completed its first pericentric passage. A Spearman rank test shows a  correlation, at the level of $2.8\sigma{}$ and $2.2\sigma{}$, respectively, between pericentric time and both the time when $f_\mathrm{gas}$ drops below $1\%$ and the time when the galaxy turns red. \label{fig:ColorTransformation}}
\end{center}
\end{figure}

\subsection{The Drivers of the Transformation Processes}
\label{sect:Drivers}

At $z\gtrsim{}1$, the most massive progenitors of the $z=0$ group satellite galaxies are blue ($B-I<1$), strongly star-forming (specific SFR $>$ $t_\mathrm{H}^{-1}$), gas-rich ($f_\mathrm{gas}\sim{}20\%$), and rotationally-supported  ($v_\mathrm{rot}/\sigma_\mathrm{cen}\geq{}1$), see Table~\ref{tab:StatGroupSat} and Figure~\ref{fig:VelSigmaMass}.
  In the simulation we can trace when, and how,  the progenitor disks of  the passively evolving spheroids and ``gas-starved'' disks, exhausted and/or expelled their  fuel for star formation, and when and how they transformed their morphologies (in those cases they did).  

\begin{figure*}
\begin{center}
\begin{tabular}{ccc}
\includegraphics[width=55mm]{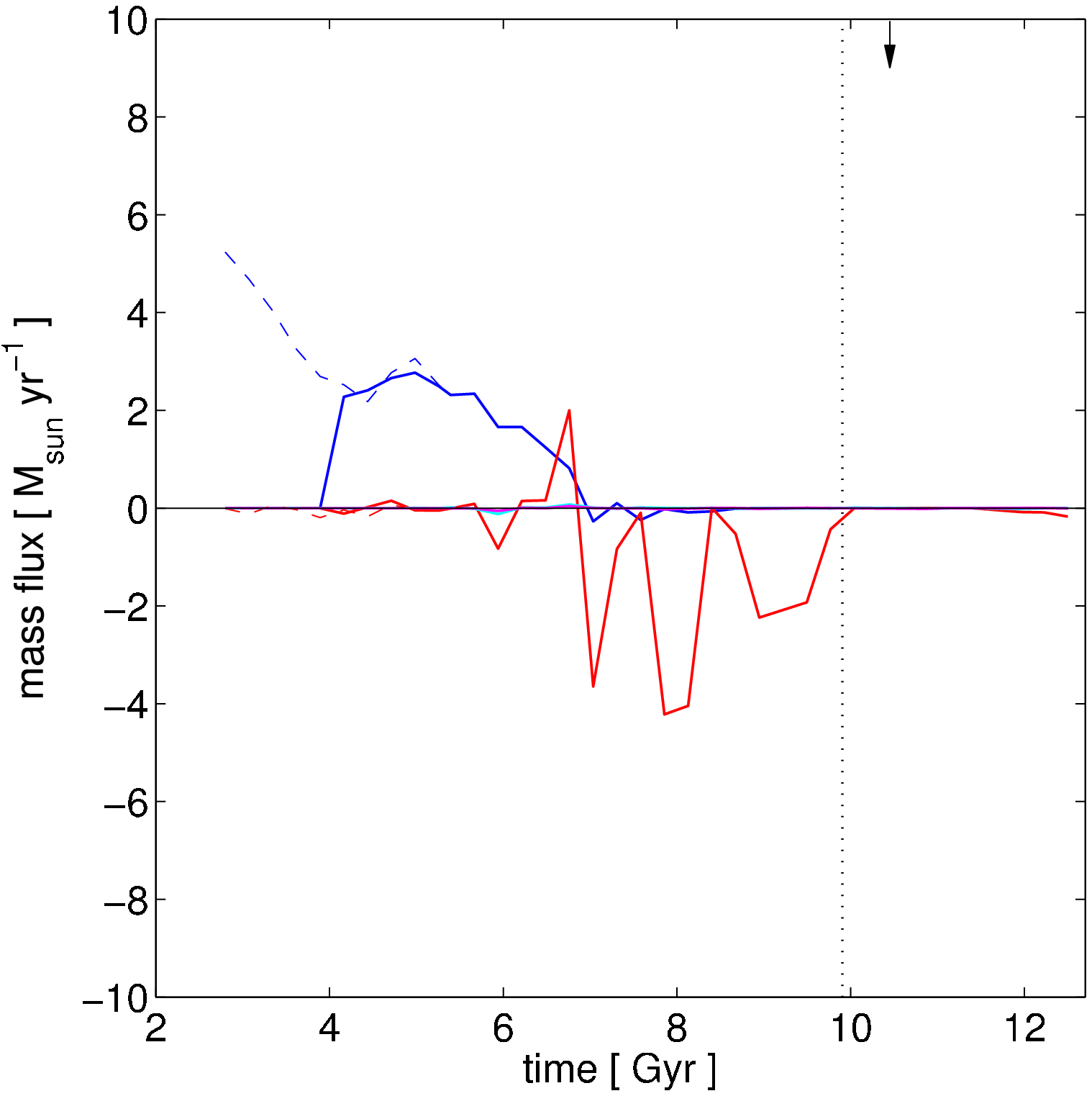} &
\includegraphics[width=55mm]{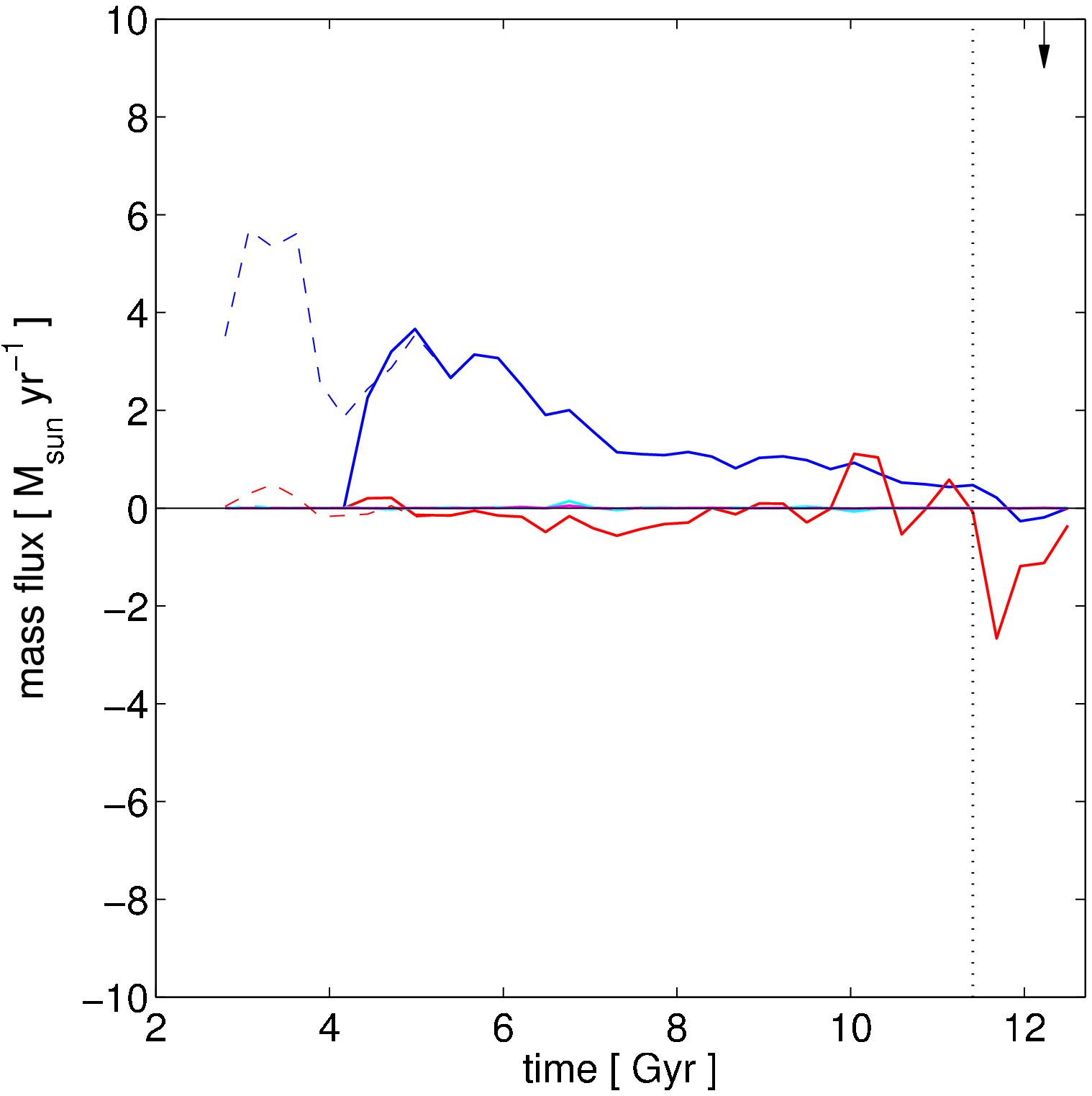} &
\includegraphics[width=55mm]{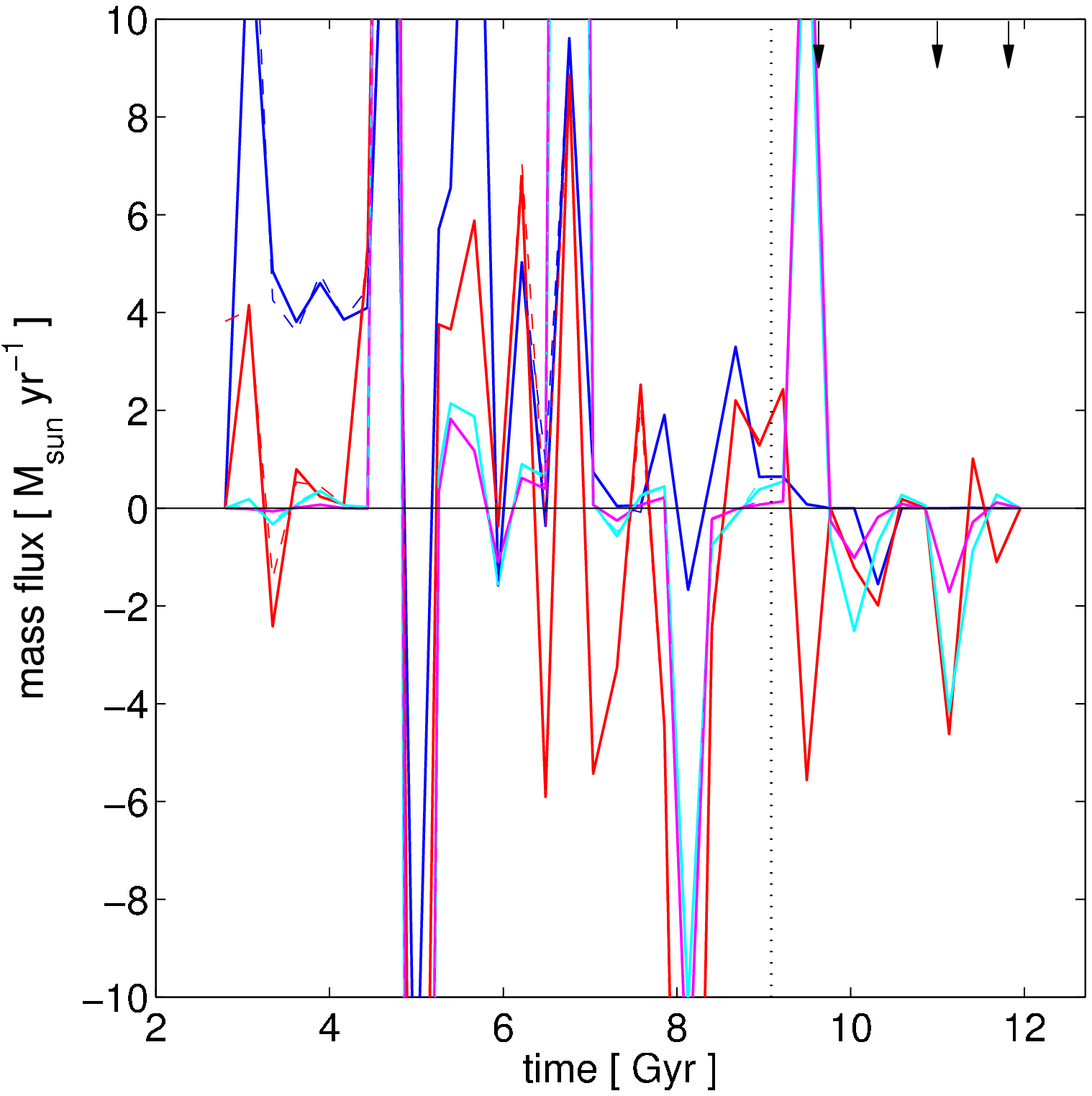} 
\end{tabular}
\caption{Gas accretion history of (left) a $z=0.1$  gas-poor disk satellite (3 in Figure 2), (middle) a $z=0.1$ intermediate-gas disk satellite (2 in Figure 2), and (right) a $z=0.1$ 
elliptical satellite (13 in Figure 2). Shown are the instantaneous accretion rates of gas that was never heated above $T_\mathrm{thres}=2.5\times{}10^{5}$ K (cold accretion; blue line), and of gas heated above that threshold (hot accretion; red line). The dotted vertical line denotes the infall time into the group, and arrows indicate subsequent pericentric passages. The solid line includes only particles bound to the respective satellite, while the dashed line includes all particles (only shown before infall). The right panel also shows the mass accretion rates of stellar particles that are formed by gas particles once heated (cyan) or not heated (magenta) above $T_\mathrm{thres}$.
The gas accretion of disk satellite progenitors (left and middle panel) proceeds in a smooth fashion, but is very irregular for the $z=0$ elliptical satellite. The gas accretion of the gas-poor disk is already quenched at $t\sim{}7$ Gyr, when it resides in a small neighboring group. The gas accretion of the intermediate-gas disk is terminated after infall to the main group. Just before the infall the density of the hot gas increases as the satellite enters the group halo, which is filled with hot, shock-heated intragroup gas. Shortly afterward, ram pressure stripping removes a large fraction of the satellite's own hot halo gas \citep{2008MNRAS.383..593M}. In the case of the elliptical satellite cold accretion dominates at early times ($t<4$ Gyr). At later times, but before the fall into the group, the gas accretion history is dominated by mergers. At the latest epochs  the group environment quenches any further gas accretion onto the satellite.
\label{fig:Acc}}
\end{center}
\end{figure*}

\begin{figure*}
\begin{center}
\begin{tabular}{ccc}
\includegraphics[width=55mm]{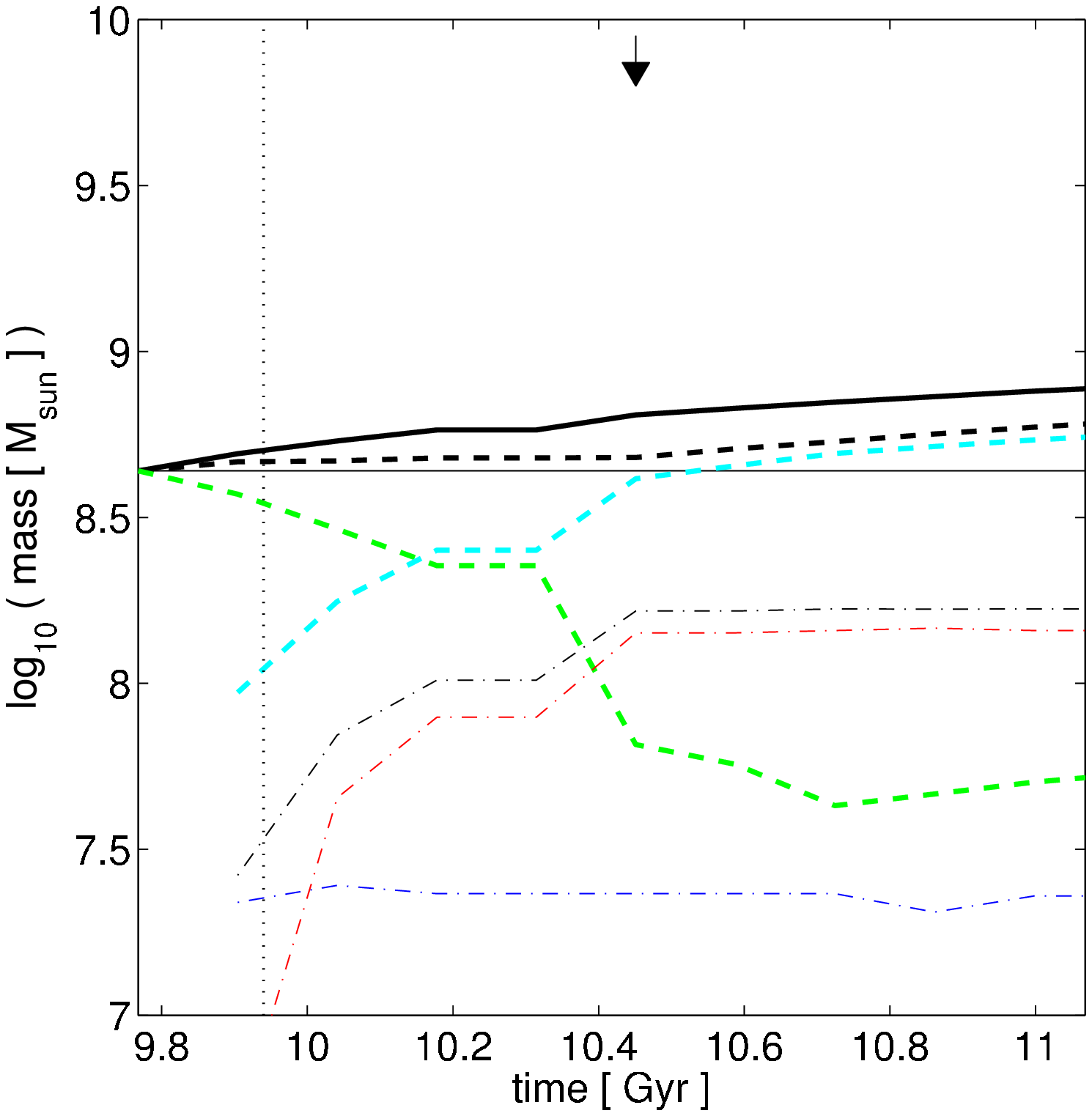} &
\includegraphics[width=55mm]{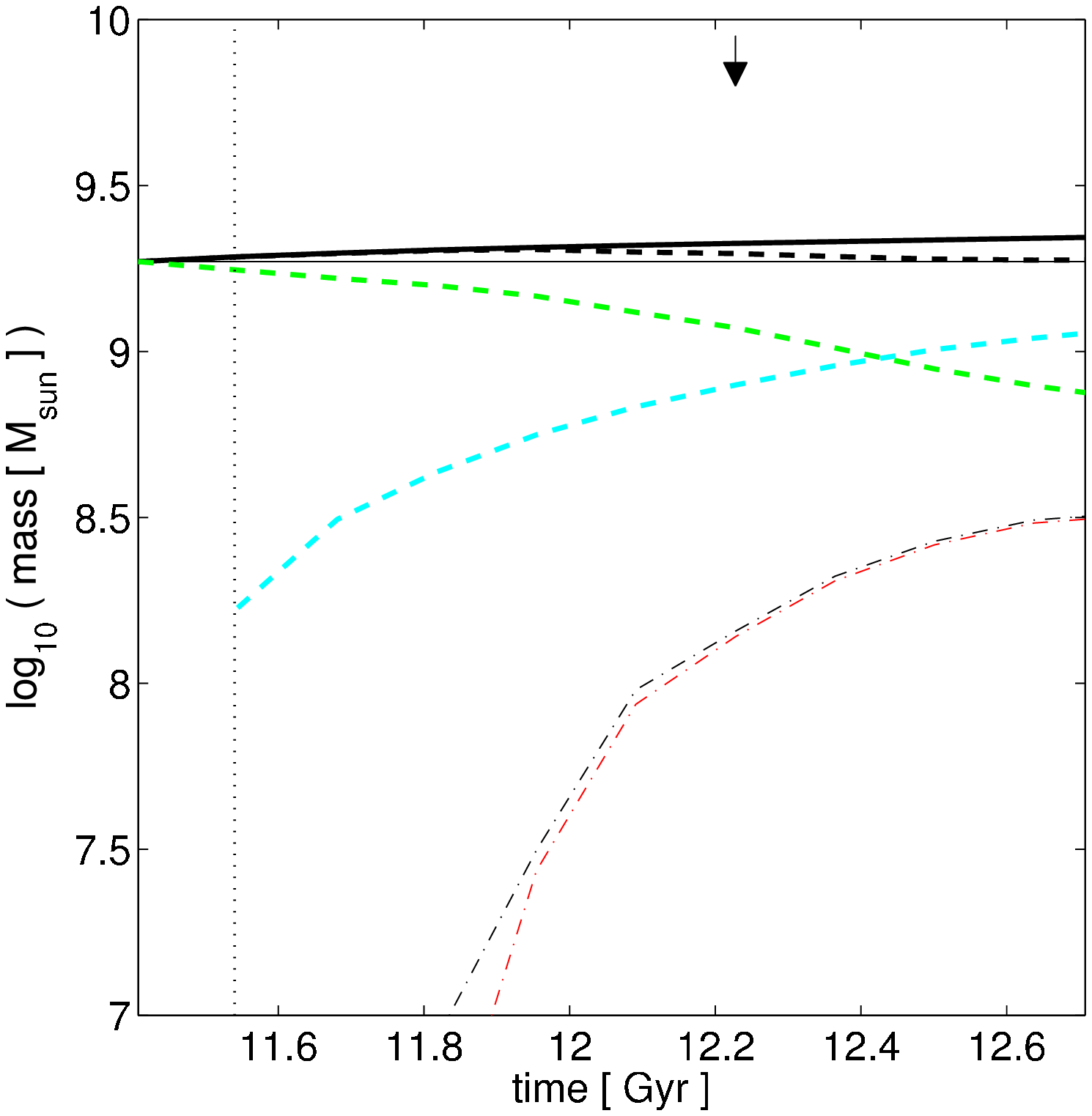} &
\includegraphics[width=55mm]{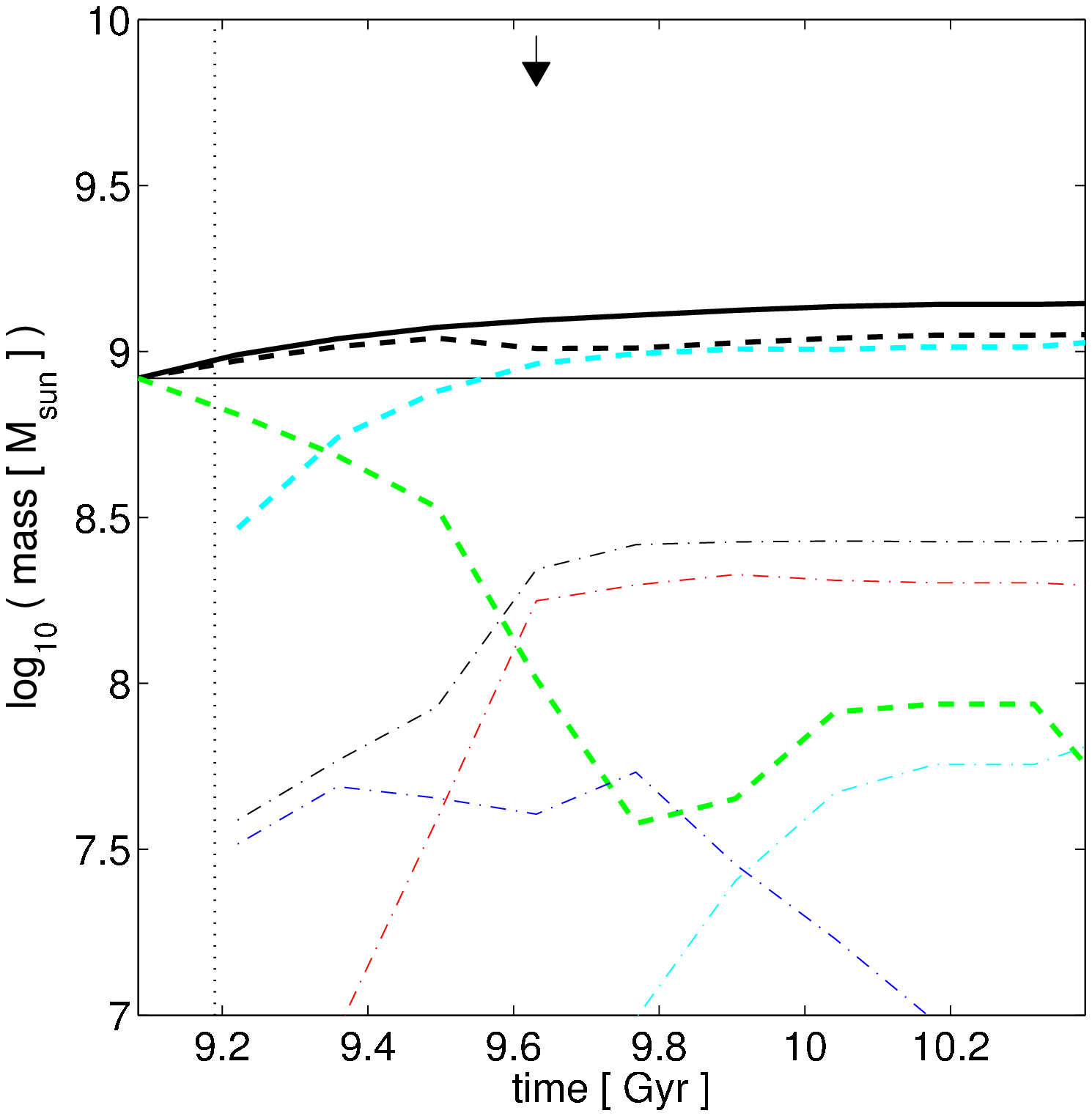} 
\end{tabular}
\caption{Fate of the cold gas in (left) a $z=0.1$ gas-poor disk satellite (3 in Figure~2), (middle) a $z=0.1$ intermediate-gas disk satellite (2 in Figure~2), and (right) a $z=0.1$ elliptical satellite (13 in Figure~2). The cold gas particles contained within a sphere of $20$ kpc radius around the satellite are identified in a snapshot before infall time (vertical dotted line) and traced forward in time. The solid black line shows the total mass of all successor particles of the selected cold gas particles. This mass is not conserved but slightly increases due to stellar winds from the satellite's own stellar population. The mass that remains within ($<20$ kpc) the satellite is shown as dashed black line, while the mass of stripped ($>20$ kpc) particles is shown with a dot-dashed black line. The other lines correspond to the amount of gas that remains and stays cold (dashed green line), that is transformed into stars that remain within the satellite (dashed cyan line), that is stripped (dot-dashed black line), that is stripped and heated (dot-dashed red line), that is stripped and stays cold (dot-dashed blue line), or that is transformed into stripped stars (dot-dashed cyan line). Most of the cold gas reservoir present before infall into the group is converted into stars within 1 Gyr or less, while a small fraction is stripped between infall time and the first pericentric passage (vertical arrow).
\label{fig:Trace}}
\end{center}
\end{figure*}


 Major galaxy mergers are in most cases responsible for the morphological transformation of disk galaxies into systems with an elliptical morphology.
Remarkably, and rather unexpectedly, however, these transformations happen before galaxies enter the virial radius of the group. They
produce six out of the seven ellipticals that are present in the group by the present epoch. Therefore, while groups have always been considered
as the natural site for major galaxy mergers given their low velocity dispersion, we find that elliptical galaxies
in groups are the result of pre-processing at $z\gtrsim{}1$, caused by interactions occurring when galaxies are still distinct objects
embedded in their individual halos. The galaxies are also still gas-rich when they undergo such mergers, and they will lose
their gas only later when they will fall into the virialized region of the group. 
Figure~\ref{fig:StellMass} shows  that  the progenitors of $z=0.1$ elliptical satellites all suffered from (moderately dissipative, $f_\mathrm{gas}\sim{}10\%-20\%$) major  ($>1:6$)  merger events over the $z\sim1-2$ period.  In contrast, the progenitors of $z=0$ disk satellites have a rather quiet and uneventful history: none of them experiences  a significant 
($>1:10$) stellar merger after redshift $z\sim{}2$. 
These galaxies  grow their stellar mass by in situ star formation, which  typically  proceeds at a rate $\sim{}1$ $M_\odot$ yr$^{-1}$ at $z<1$.  
The mergers that produce the elliptical morphologies naturally lead  to a faster growth of stellar mass relative to in situ star formation.

\begin{figure}
\begin{center}
\begin{tabular}{c}
\includegraphics[width=83mm]{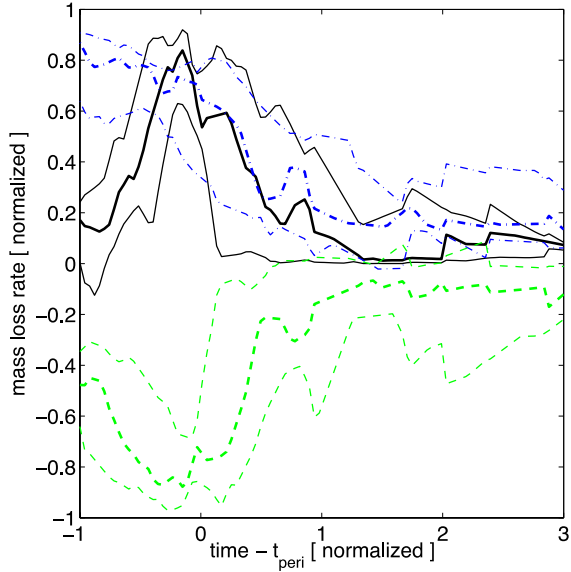}
\end{tabular}
\caption{Effects of pericenter passages in the evolution of the satellites. The figure shows the average cold gas transformation rates obtained after stacking all satellites together on normalized axes. The time axis is normalized in units of $t_\mathrm{peri}-t_\mathrm{infall}$, while the rates of each individual satellite are normalized to its maximum value. Thick (thin) lines denote the median (lower/upper quartile) of the satellite population.
Most of the cold gas that resides in satellites just before they infall into the group (green dashed lines; shown with the negative of the mass loss rate for better visualization) is either continuously  transformed into stars (blue dot-dashed lines) or stripped (black solid lines) in pericentric passages. Some satellites show enhanced star formation rates close to pericenter time. However, overall, the conversion of cold gas into stars is strongly reduced after the first pericentric passage. Within a timescale of about $t_\mathrm{peri}-t_\mathrm{infall}$ after pericenter passage, the cold gas reservoir is exhausted and the mass loss rates become negligible.\label{fig:RoleOfPericenter}}
\end{center}
\end{figure}

\begin{figure}
\begin{center}
\includegraphics[width=85mm]{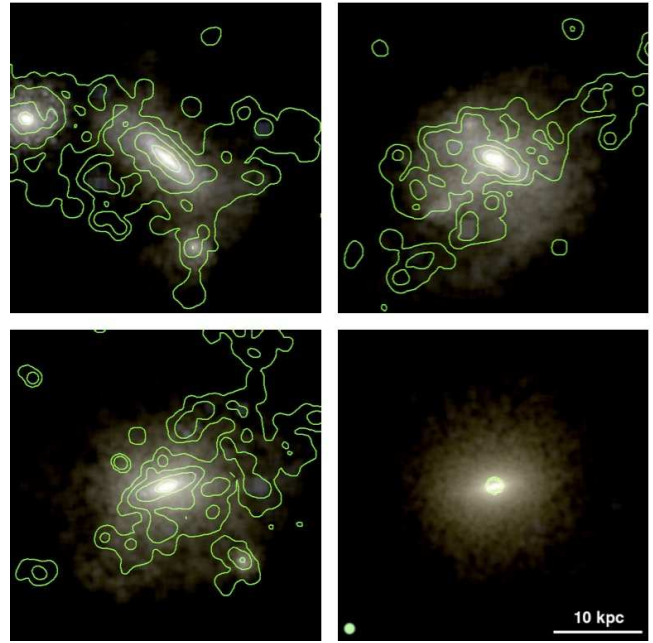}
\caption{Formation of a $z=0.1$ elliptical galaxy that lives today in  the relatively dense potential of a $\sim10^{13}M_\odot$ galaxy group. (Top left) Edge-on view of the $z=1.1$ main progenitor of the elliptical galaxy 13 in Figure~\ref{fig:GroupCensus}; the progenitor has both a stellar and a gas disk.  (Top right) A gas-rich merger occurs $\sim{}400$ Myr later, destroying the stellar disk. (Bottom left) Another 300 Myr later the stellar component of the galaxy has relaxed to an elliptical morphology while still showing blue colors. Also, a small gas disk begins to grow again, out of the remaining cold gas. (Bottom right)  At $z=0.38$ the galaxy completes its first pericentric passage. By this time the galaxy has lost most of its cold gas, its star formation is quenched, and the colors of the galaxy are now typical of  ``red sequence'', passively evolving galaxies. 
Column densities of neutral hydrogen are indicated by green contour lines which correspond to 0.1, 1, 10, and 100 $M_\odot$ pc$^{-2}$, respectively.
\label{fig:Merger}}
\end{center}
\end{figure}

The group potential plays on the other hand a crucial role in removing the gas reservoir of all galaxies and truncating their star formation (Figure \ref{fig:ColorTransformation}). 
As a result, the photometric evolution of ellipticals into the passively evolving early-type galaxy population observed at the present epoch lags behind by several Gyr the time
at which their spheroidal morphology is established in the mergers at earlier epochs. Likewise, the transformation of normal star-forming disks into gas-poor, ``starved''  red disks  
appears to be  mostly a consequence of their infall into group environment.
Indeed, the three disk galaxies that are  gas-poor and quiescent by $z=0$  enter the group at earlier epochs than the three $z=0$ star-forming disks
(respectively, $z=0.35,0.49,0.80$ and $z=0.13,0.19,0.26$), and are exposed for longer periods to the physical processes that are most effective in the group environment. 
Such processes are, precisely, suppression of gas accretion as well as ram pressure stripping of both the hot and cold gas reservoir; in situ star formation is responsible for consuming
a substantial fraction of the residual gas.
In detail, a galaxy that plunges into the depth of the group potential is first observed to stop accreting gas after the infall. Subsequently, a substantial amount of its hot halo gas is 
removed by ram-pressure stripping on its first orbit within the group, namely on a timescale of a Gyr or so, consistent with a ``starvation'' picture \citep{1980ApJ...237..692L, 2008ApJ...672L.103K}, see Figure~\ref{fig:Acc}. Most of the cold gas ($\sim{}90\pm{}10\%$)  is then  lost within the following $\sim{}1$ Gyr, through star formation ($\sim{}70\%$) as well as nearly instantaneous gas stripping at pericenter passages (Figures~\ref{fig:Trace}, and \ref{fig:RoleOfPericenter}). For galaxies falling into the final virialized group halo, the first pericenter passage is reached $0.4\pm{}0.1$ Gyr after infall into the group, i.e., the characteristic crossing time of the group (Figure \ref{fig:ColorTransformation}).

\begin{figure*}
\begin{center}
\begin{tabular}{ccc}
\includegraphics[width=52mm]{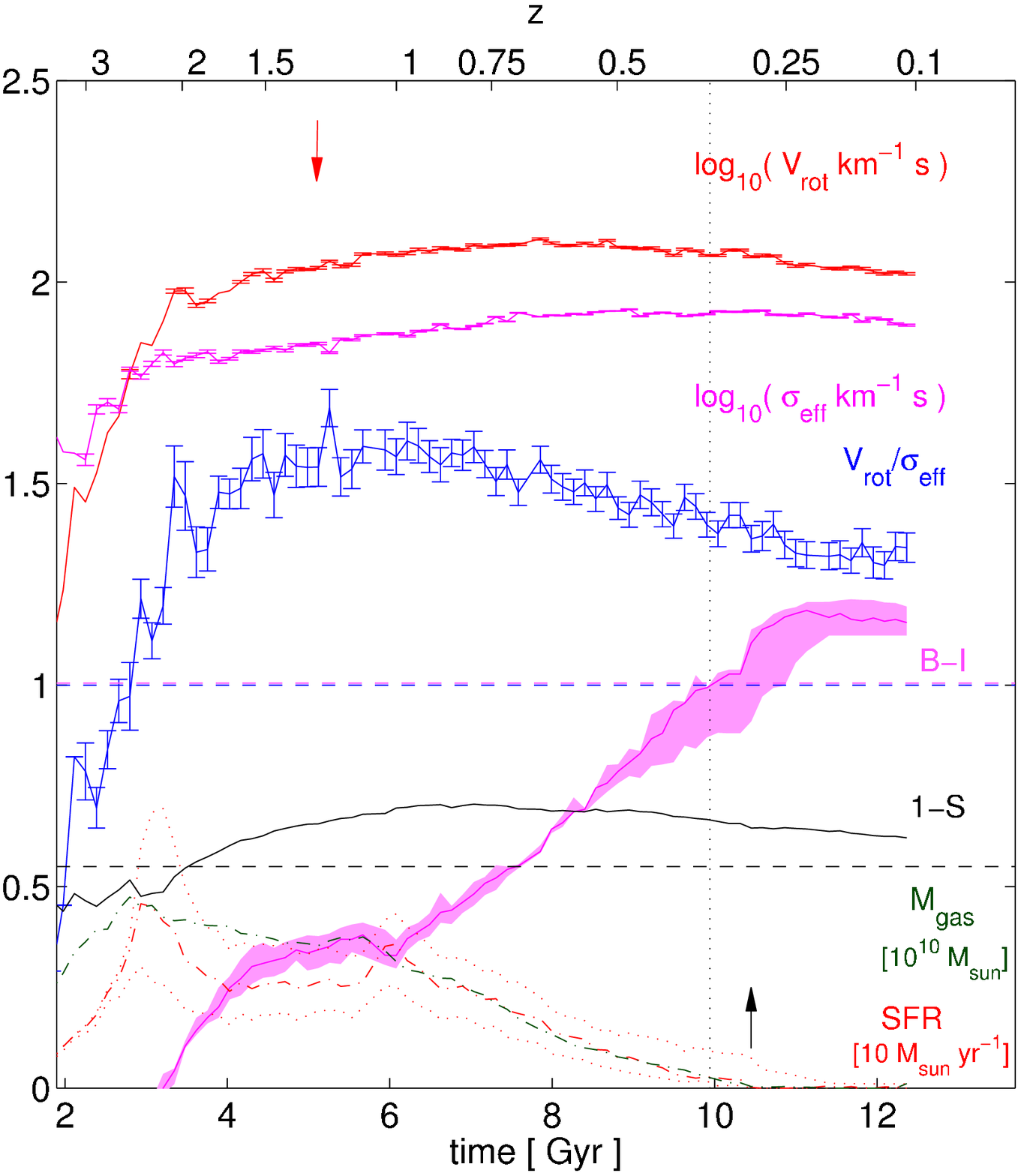} &
\includegraphics[width=52mm]{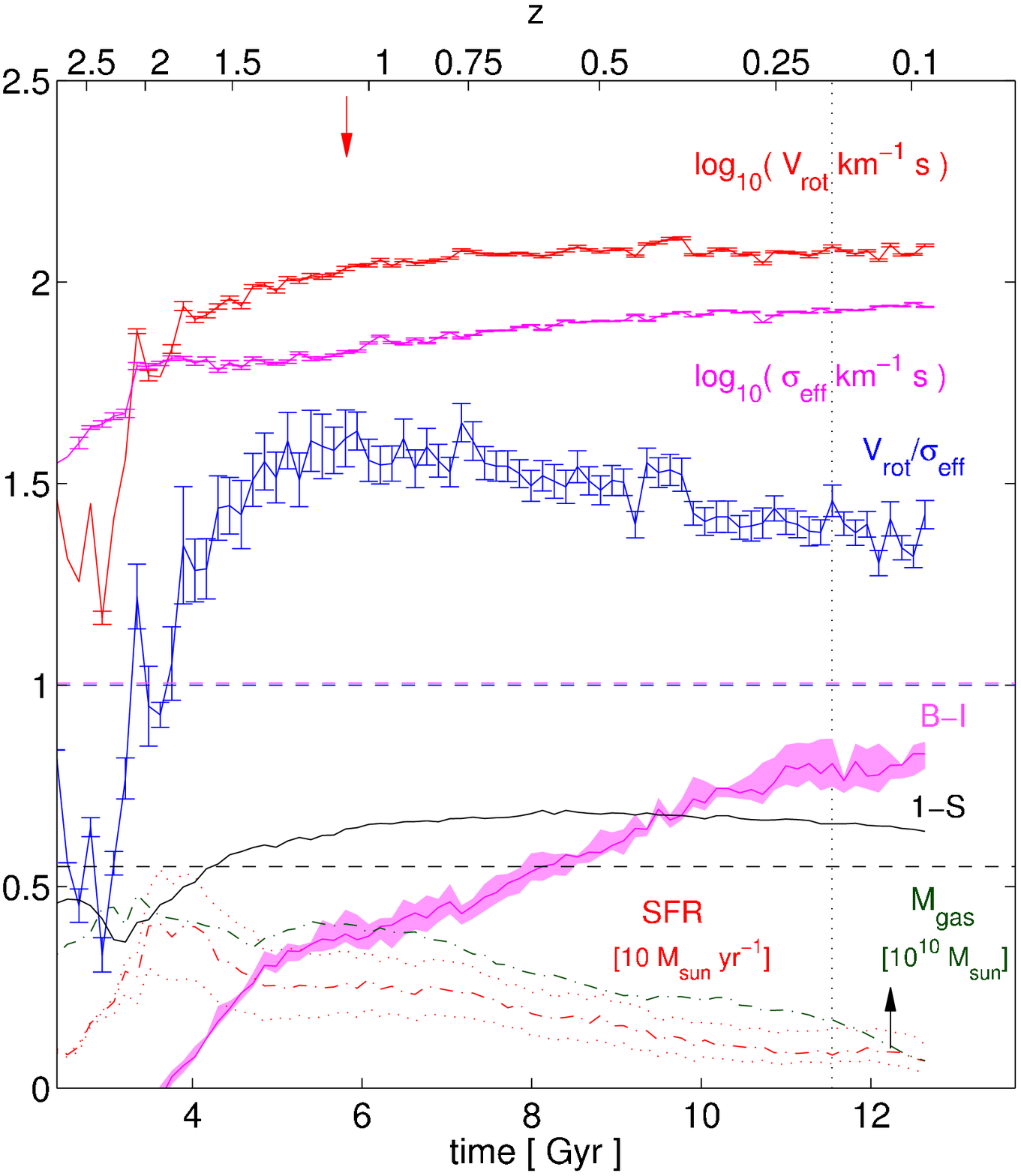} &
\includegraphics[width=52mm]{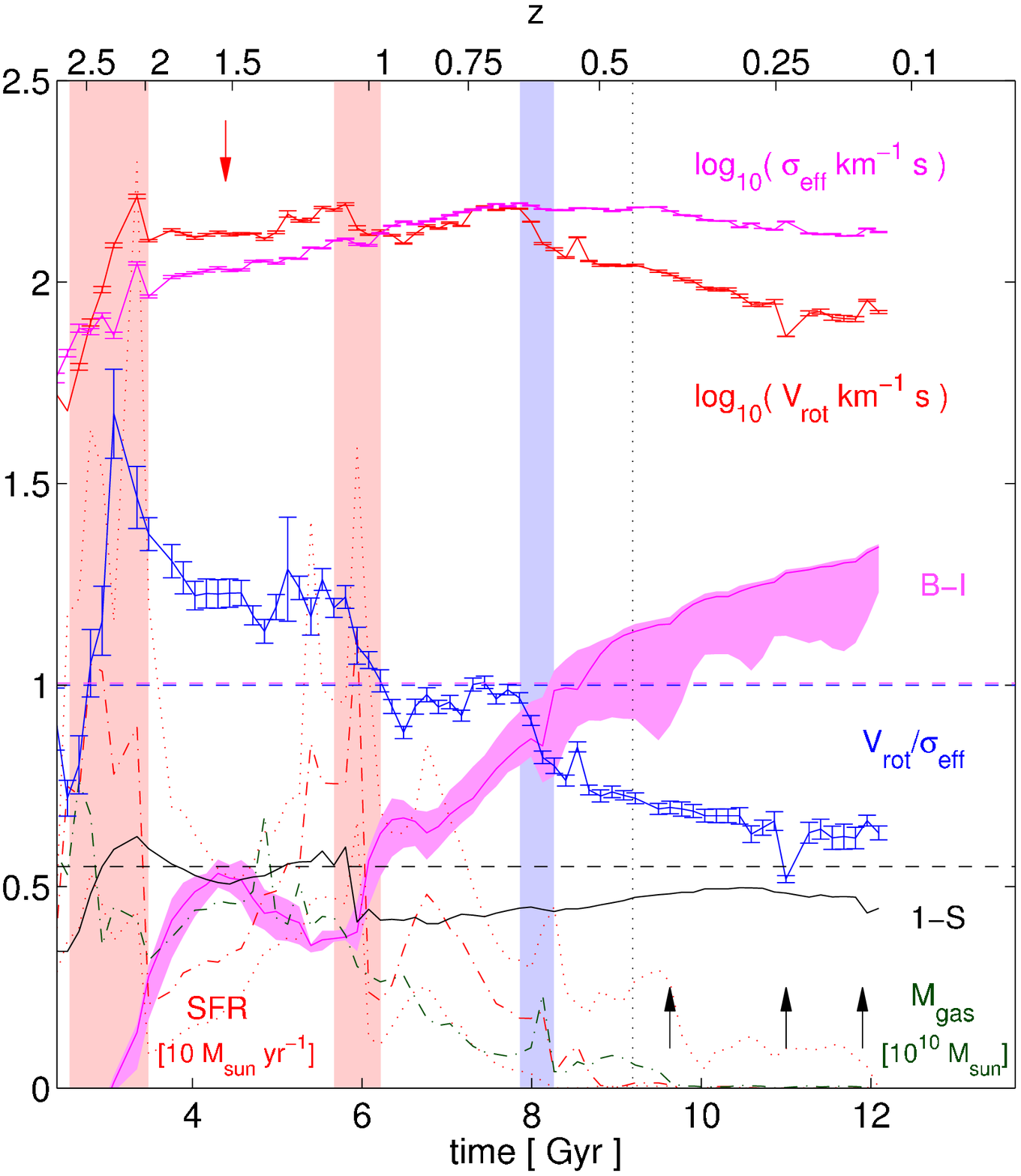}
\end{tabular}
\caption{Evolution of morphological, kinematic, and photometric properties of (left) a $z=0.1$ gas-poor disk satellite (3 in Figure~2), (middle) a $z=0.1$ intermediate-gas disk satellite (2 in Figure~2), and (right) a $z=0.1$ elliptical satellite (13 in Figure~2). Shown are the line-of-sight velocity dispersion (magenta line with error bars), the maximum of the line-of-sight rotation velocity (red line with error bars), $v/\sigma$ (blue line with error bars), the non-sphericity 1- $c/a$ (black solid line), the amount of cold gas ($<3.2\times{}10^{4}$ K) within 20 kpc (green dot-dashed line), the star formation rate within 10 kpc (upper red dotted line), within 10 kpc excluding one softening length (red dot-dashed line), within 10 kpc excluding two softening lengths (lower red dotted line), and the $B-I$ color (magenta shaded area; the upper and lower limits are measured within 10 kpc excluding and including the central two softening lengths, while the magenta line is measured by excluding the central softening length). The red arrow at the top left is the time when half of the stellar mass is formed. The vertical dotted line denotes the time of infall into the group and the black arrows on the bottom right denote pericentric passages. The vertical shaded areas in the right panel denote prominent mergers (red) or a fly-by (blue). 
\label{fig:TimeEvolution}}
\end{center}
\end{figure*}

In Figure~\ref{fig:Merger} we illustrate, as an example,  how the  transformation into a gas-poor, quiescent, red, velocity-dispersion supported elliptical galaxy takes place as a two-step process, starting  from gas-rich, star-forming, blue, rotating galaxies. First, at $z\sim{}1.1$ and at $\sim{}1$ Mpc distance from the group center, a gas-rich merger with a stellar mass ratio $\sim{}1:2$ destroys the stellar disk of the more massive progenitor. The remnant relaxes in a few hundred Myr to an elliptical morphology,  but its colors remain blue and it still hosts a substantial amount ($\sim{}10^9$ $M_\odot$) of HI within 20 kpc. During the merger the SFR is substantially enhanced \citep{1996ApJ...464..641M}; at its peak,  it is about a factor five higher than the average rate that the galaxy experiences throughout its history. Thereafter the HI reservoir and the SFR gradually decrease. After the first pericenter, the SFR and the HI content become negligible and the galaxy turns into a red, passively evolving elliptical ($\sim L*$ in the SDSS $z=0.1$ redshifted $r$-filter; \citealt{2003ApJ...592..819B}) with a final $B-I=1.34$. 

We note  that one of the seven $z=0$ elliptical satellites does not undergo any substantial merger since $z\sim{}2.2$ (red triangle in figure Figure~\ref{fig:StellMass}). This galaxy assembles most of its stellar mass early on,  i.e., much earlier than most disk satellites. Not surprisingly, this results in a massive and very compact spheroidal system. We speculate that the very diverse stellar density and velocity dispersion properties observed in the $z\sim2$  passive galaxy population \citep{2009Natur.460..717V} 
could be, at least in part, explained by a spread in the assembly epochs of its constituent galaxies.

\subsection{Environmental Effects before Infall into the Group}
\label{sect:Environment}

Although the ultimate quenching of galaxies is caused by stripping of the gas reservoir and the resulting suppression of star formation
after galaxies enter the group potential, we note that their star formation histories start being affected by environmental
effects also prior to their infall. For most galaxies,
the SFR peaks around $z\sim{}2$ at a distance of $\sim{}1$ Mpc ($\sim{}$turn around radius) from the center of the assembling group.
Afterward, a different behavior is seen in the progenitors of the present-day population of, respectively, disk and elliptical galaxies.
In the progenitors of disks, star formation declines smoothly from high to low redshift until truncation ensues, reflecting the fact
that these galaxies evolve in near isolation before joining the group. In contrast, the SFR of the progenitors of ellipticals, which are gas-rich,
massive disk-like objects,  shows high variability, with bursts caused by galaxy interactions and mergers (Figure \ref{fig:TimeEvolution}).

As similar dichotomy occurs for gas accretion, which happens predominantly in the ``cold mode'' for all galaxies prior to infall onto the group.
In galaxies which remain disks down to $z=0$,  the accretion rate decreases gradually with time \citep{2003ApJ...596...47S}, see Figure \ref{fig:Acc}.
In the early-type galaxies, the rate is instead rather bursty and strongly modulated by the mergers these systems
endure. In some galaxies, the decrease in the cold accretion rate  is exacerbated by ``truncation'' of the streams
due to the reduction of the tidal radius of the satellites (driven by the tidal  effects of the group halo potential; \citealt{2009MNRAS.tmp.1114H}).
This leads to quenching of gas accretion already at a few virial radii away from the group center.  In at least one of the gas-poor disks (Figures~\ref{fig:Acc}, and \ref{fig:TimeEvolution}),  the supply of cold gas is already terminated when it resides  in the vicinity of a $\sim{}5\times{}10^{11} M_\odot$ halo
that falls into the main group at a substantially later time.

\section{Conclusions}
\label{sect:Conclusions}

Our work shows that the formation of $\lesssim{}L*$ elliptical galaxies and ``starved'' red disk galaxies in high-density environments 
such as galaxy groups is the result of the combined effect of merging, gas removal and gas consumption processes as well as halted gas accretion naturally
occurring as the potential of the group is assembled hierarchically in a $\Lambda{}CDM$
universe. Our simulations do not include feedback from AGNs and thus show that it is not an essential ingredient in order to form elliptical galaxies with quenched star formation in $z\sim{}0$ galaxy groups.

Specifically, we find that elliptical galaxies are formed by mergers at high redshift \emph{prior} to the formation of the final group, which is consistent with the fact that elliptical galaxies are found to be already in place at high redshift even in the field \citep{2006A&A...453L..29C, 2007ApJS..172..494S, 2009arXiv0907.0013B}.  
The lag of photometric transformations relative to the morphological transformations seen in our simulation indicates that many of the  ``excess'' blue $L*$ galaxies 
observed at $z\sim1$,  which are thought to become $L*$ ellipticals, but which appear to have an irregular morphology at blue rest-frame wavelengths \citep{2007ApJS..172..494S, 2009arXiv0907.0013B}, should show  an elliptical morphology when observed at longer rest-frame wavelengths. This is a straight prediction of  our simulations since at the time when their old stars
acquire an elliptical morphology through mergers, the simulated galaxies still host a star-forming, gaseous component. This also indicates that the tight correlation of colors and Hubble types, as well as the standard classification based on the Hubble sequence, does not hold for galaxies at redshifts higher than about $1.5-2$, in good agreement with  state-of-the-art studies of galaxy morphologies at such high redshifts \citep{2010arXiv1007.2422C}.
%
%

In addition, our model predicts that the star formation and gas accretion rates of progenitors of elliptical galaxies in the considered mass range ($\sim{}3\times{}10^{10} M_\odot$) should be highly irregular due to gas-rich galaxy interactions and mergers at intermediate redshifts $z\gtrsim{}1$. Therefore, these galaxies should show a large scatter
in the ages of their stellar populations and differ from single, high redshift burst models \citep{2005MNRAS.362...41G, 2005ApJ...621..673T}. Future multi-wavelength observations capable of characterizing with unprecedented detail both the SFR evolution and the evolution of the gas content of galaxies, atomic and molecular, as a function of redshift, such as the Atacama Large Millimeter Array, the James Webb Space Telescope, and the Square Kilometer Array, will be able to test this scenario.



\paragraph{Acknowledgments}
R.F. acknowledges funding by the Swiss National Science Foundation. The simulation has been carried out at the Swiss National Computing Center (CSCS in Manno).



\bibliographystyle{apj}

\begin{thebibliography}{56}
\expandafter\ifx\csname natexlab\endcsname\relax\def\natexlab#1{#1}\fi

\bibitem[Abadi et al.(2003)]{2003ApJ...591..499A} Abadi, M.~G., Navarro, 
J.~F., Steinmetz, M., \& Eke, V.~R.\ 2003, \apj, 591, 499 

\bibitem[Agertz et al.(2011)]{2011MNRAS.410.1391A} Agertz, O., Teyssier, 
R., \& Moore, B.\ 2011, \mnras, 410, 1391 

\bibitem[{{Balogh} \& {Morris}(2000)}]{2000MNRAS.318..703B}
{Balogh}, M.~L., \& {Morris}, S.~L. 2000, \mnras, 318, 703

\bibitem[{{Bell} \& {de Jong}(2001)}]{2001ApJ...550..212B}
{Bell}, E.~F., \& {de Jong}, R.~S. 2001, \apj, 550, 212

\bibitem[{{Bender} {et~al.}(1992){Bender}, {Burstein}, \&
  {Faber}}]{1992ApJ...399..462B}
{Bender}, R., {Burstein}, D., \& {Faber}, S.~M. 1992, \apj, 399, 462

\bibitem[{{Benson} \& {Devereux}(2010)}]{2010MNRAS.402.2321B}
{Benson}, A.~J., \& {Devereux}, N. 2010, \mnras, 402, 2321

\bibitem[{{Bertschinger}(2001)}]{2001ApJS..137....1B}
{Bertschinger}, E. 2001, \apjs, 137, 1

\bibitem[{{Bessell}(1990)}]{1990PASP..102.1181B}
{Bessell}, M.~S. 1990, \pasp, 102, 1181

\bibitem[{{Birnboim} \& {Dekel}(2003)}]{2003MNRAS.345..349B}
{Birnboim}, Y., \& {Dekel}, A. 2003, \mnras, 345, 349

\bibitem[{{Blanton} {et~al.}(2003){Blanton}, {Hogg}, {Bahcall}, {Brinkmann},
  {Britton}, {Connolly}, {Csabai}, {Fukugita}, {Loveday}, {Meiksin}, {Munn},
  {Nichol}, {Okamura}, {Quinn}, {Schneider}, {Shimasaku}, {Strauss}, {Tegmark},
  {Vogeley}, \& {Weinberg}}]{2003ApJ...592..819B}
{Blanton}, M.~R., et al. 2003, \apj, 592, 819

\bibitem[Bolzonella et 
al.(2010)]{2009arXiv0907.0013B} Bolzonella, M., et al.\ 2010, \aap, 524, A76 

\bibitem[{{Brooks} {et~al.}(2009){Brooks}, {Governato}, {Quinn}, {Brook}, \&
  {Wadsley}}]{2009ApJ...694..396B}
{Brooks}, A.~M., {Governato}, F., {Quinn}, T., {Brook}, C.~B., \& {Wadsley}, J.
  2009, \apj, 694, 396

\bibitem[{{Bruzual} \& {Charlot}(2003)}]{2003MNRAS.344.1000B}
{Bruzual}, G., \& {Charlot}, S. 2003, \mnras, 344, 1000

\bibitem[Cameron et al.(2010)]{2010arXiv1007.2422C} Cameron, E., Carollo, 
C.~M., Oesch, P.~A., Bouwens, R.~J., Illingworth, G.~D., Trenti, M., Labbe, 
I., \& Magee, D.\ 2010, arXiv:1007.2422 

\bibitem[{{Carollo} {et~al.}(1993){Carollo}, {Danziger}, \&
  {Buson}}]{1993MNRAS.265..553C}
{Carollo}, C.~M., {Danziger}, I.~J., \& {Buson}, L. 1993, \mnras, 265, 553

\bibitem[{{Charlot} \& {Fall}(2000)}]{2000ApJ...539..718C}
{Charlot}, S., \& {Fall}, S.~M. 2000, \apj, 539, 718

\bibitem[{{Cimatti} {et~al.}(2006){Cimatti}, {Daddi}, \&
  {Renzini}}]{2006A&A...453L..29C}
{Cimatti}, A., {Daddi}, E., \& {Renzini}, A. 2006, \aap, 453, L29

\bibitem[{{Cox} {et~al.}(2006){Cox}, {Dutta}, {Di Matteo}, {Hernquist},
  {Hopkins}, {Robertson}, \& {Springel}}]{2006ApJ...650..791C}
{Cox}, T.~J., {Dutta}, S.~N., {Di Matteo}, T., {Hernquist}, L., {Hopkins},
  P.~F., {Robertson}, B., \& {Springel}, V. 2006, \apj, 650, 791

\bibitem[{{Croft} {et~al.}(2009){Croft}, {Di Matteo}, {Springel}, \&
  {Hernquist}}]{2009MNRAS.400...43C}
{Croft}, R.~A.~C., {Di Matteo}, T., {Springel}, V., \& {Hernquist}, L. 2009,
  \mnras, 400, 43

\bibitem[Dav{\'e}(2009)]{2009ASPC..419..347D} Dav{\'e}, R.\ 2009, 
in ASP Conf. Ser 419, Galaxy Evolution: Emerging Insights and Future Challenges, ed. S. Jogee, I. Marinova, L. Hao, \& G. A. Blanc (San Francisco, CA: ASP), 347

\bibitem[{{Dressler} {et~al.}(1987){Dressler}, {Lynden-Bell}, {Burstein},
  {Davies}, {Faber}, {Terlevich}, \& {Wegner}}]{1987ApJ...313...42D}
{Dressler}, A., {Lynden-Bell}, D., {Burstein}, D., {Davies}, R.~L., {Faber},
  S.~M., {Terlevich}, R., \& {Wegner}, G. 1987, \apj, 313, 42

\bibitem[{{Feldmann} {et~al.}(2010){Feldmann}, {Carollo}, {Mayer}, {Renzini},
  {Lake}, {Quinn}, {Stinson}, \& {Yepes}}]{2010ApJ...709..218F}
{Feldmann}, R., {Carollo}, C.~M., {Mayer}, L., {Renzini}, A., {Lake}, G.,
  {Quinn}, T., {Stinson}, G.~S., \& {Yepes}, G. 2010, \apj, 709, 218

\bibitem[Fukugita et al.(1996)]{1996AJ....111.1748F} Fukugita, M., 
Ichikawa, T., Gunn, J.~E., Doi, M., Shimasaku, K., 
\& Schneider, D.~P.\ 1996, \aj, 111, 1748 

\bibitem[{{Gallazzi} {et~al.}(2005){Gallazzi}, {Charlot}, {Brinchmann},
  {White}, \& {Tremonti}}]{2005MNRAS.362...41G}
{Gallazzi}, A., {Charlot}, S., {Brinchmann}, J., {White}, S.~D.~M., \&
  {Tremonti}, C.~A. 2005, \mnras, 362, 41

\bibitem[{{Gill} {et~al.}(2004){Gill}, {Knebe}, \&
  {Gibson}}]{2004MNRAS.351..399G}
{Gill}, S.~P.~D., {Knebe}, A., \& {Gibson}, B.~K. 2004, \mnras, 351, 399

\bibitem[{{Governato} {et~al.}(2007){Governato}, {Willman}, {Mayer}, {Brooks},
  {Stinson}, {Valenzuela}, {Wadsley}, \& {Quinn}}]{2007MNRAS.374.1479G}
{Governato}, F., {Willman}, B., {Mayer}, L., {Brooks}, A., {Stinson}, G.,
  {Valenzuela}, O., {Wadsley}, J., \& {Quinn}, T. 2007, \mnras, 374, 1479

\bibitem[Governato et al.(2010)]{2010Natur.463..203G} Governato, F., et 
al.\ 2010, \nat, 463, 203

\bibitem[{{Graves} {et~al.}(2009){Graves}, {Faber}, \&
  {Schiavon}}]{2009ApJ...698.1590G}
{Graves}, G.~J., {Faber}, S.~M., \& {Schiavon}, R.~P. 2009, \apj, 698, 1590

\bibitem[Guedes et al.(2011)]{2011arXiv1103.6030G} Guedes, J., Callegari, 
S., Madau, P., \& Mayer, L.\ 2011, arXiv:1103.6030 

\bibitem[{{Gunn} \& {Gott}(1972)}]{1972ApJ...176....1G}
{Gunn}, J.~E., \& {Gott}, J.~R.~I. 1972, \apj, 176, 1

\bibitem[Guo et al.(2010)]{2010MNRAS.404.1111G} Guo, Q., White, S., Li, C., 
\& Boylan-Kolchin, M.\ 2010, \mnras, 404, 1111 

\bibitem[Haardt 
\& Madau(1996)]{1996ApJ...461...20H} Haardt, F., \& Madau, P.\ 1996, \apj, 461, 20 

\bibitem[{{Hahn} {et~al.}(2009){Hahn}, {Porciani}, {Dekel}, \&
  {Carollo}}]{2009MNRAS.tmp.1114H}
{Hahn}, O., {Porciani}, C., {Dekel}, A., \& {Carollo}, C.~M. 2009, \mnras, 398,
  1742
  
\bibitem[Hopkins et al.(2011)]{2011arXiv1101.4940H} Hopkins, P.~F., 
Quataert, E., \& Murray, N.\ 2011, arXiv:1101.4940 

\bibitem[{{Hopkins} {et~al.}(2010{\natexlab{a}}){Hopkins}, {Bundy},
  {Hernquist}, {Wuyts}, \& {Cox}}]{2010MNRAS.401.1099H}
{Hopkins}, P.~F., {Bundy}, K., {Hernquist}, L., {Wuyts}, S., \& {Cox}, T.~J.
  2010{\natexlab{a}}, \mnras, 401, 1099

\bibitem[{{Hopkins} {et~al.}(2010{\natexlab{b}}){Hopkins}, {Younger},
  {Hayward}, {Narayanan}, \& {Hernquist}}]{2010MNRAS.402.1693H}
{Hopkins}, P.~F., {Younger}, J.~D., {Hayward}, C.~C., {Narayanan}, D., \&
  {Hernquist}, L. 2010{\natexlab{b}}, \mnras, 402, 1693

\bibitem[{{Hopkins} {et~al.}(2009){Hopkins}, {Somerville}, {Cox}, {Hernquist},
  {Jogee}, {Kere{\v s}}, {Ma}, {Robertson}, \& {Stewart}}]{2009MNRAS.397..802H}
{Hopkins}, P.~F., et al. 2009,
  \mnras, 397, 802

\bibitem[{{Kawata} \& {Mulchaey}(2008)}]{2008ApJ...672L.103K}
{Kawata}, D., \& {Mulchaey}, J.~S. 2008, \apjl, 672, L103

\bibitem[{{Kere{\v s}} {et~al.}(2009){Kere{\v s}}, {Katz}, {Fardal},
  {Dav{\'e}}, \& {Weinberg}}]{2009MNRAS.395..160K}
{Kere{\v s}}, D., {Katz}, N., {Fardal}, M., {Dav{\'e}}, R., \& {Weinberg},
  D.~H. 2009, \mnras, 395, 160

\bibitem[{{Kere{\v s}} {et~al.}(2005){Kere{\v s}}, {Katz}, {Weinberg}, \&
  {Dav{\'e}}}]{2005MNRAS.363....2K}
{Kere{\v s}}, D., {Katz}, N., {Weinberg}, D.~H., \& {Dav{\'e}}, R. 2005,
  \mnras, 363, 2

\bibitem[{{Knollmann} \& {Knebe}(2009)}]{2009ApJS..182..608K}
{Knollmann}, S.~R., \& {Knebe}, A. 2009, \apjs, 182, 608

\bibitem[Kova{\v c} et al.(2010)]{2009arXiv0909.2032K} Kova{\v c}, K., et 
al.\ 2010, \apj, 718, 86 

\bibitem[{{Larson} {et~al.}(1980){Larson}, {Tinsley}, \&
  {Caldwell}}]{1980ApJ...237..692L}
{Larson}, R.~B., {Tinsley}, B.~M., \& {Caldwell}, C.~N. 1980, \apj, 237, 692

\bibitem[{{Maraston}(2005)}]{2005MNRAS.362..799M}
{Maraston}, C. 2005, \mnras, 362, 799

\bibitem[{{Mayer} {et~al.}(2008){Mayer}, {Governato}, \&
  {Kaufmann}}]{2008ASL.....1....7M}
{Mayer}, L., {Governato}, F., \& {Kaufmann}, T. 2008, Advanced Science Letters,
  1, 7

\bibitem[{{McCarthy} {et~al.}(2008){McCarthy}, {Frenk}, {Font}, {Lacey},
  {Bower}, {Mitchell}, {Balogh}, \& {Theuns}}]{2008MNRAS.383..593M}
{McCarthy}, I.~G., {Frenk}, C.~S., {Font}, A.~S., {Lacey}, C.~G., {Bower},
  R.~G., {Mitchell}, N.~L., {Balogh}, M.~L., \& {Theuns}, T. 2008, \mnras, 383,
  593

\bibitem[{{Meza} {et~al.}(2003){Meza}, {Navarro}, {Steinmetz}, \&
  {Eke}}]{2003ApJ...590..619M}
{Meza}, A., {Navarro}, J.~F., {Steinmetz}, M., \& {Eke}, V.~R. 2003, \apj, 590,
  619

\bibitem[{{Mihos} \& {Hernquist}(1996)}]{1996ApJ...464..641M}
{Mihos}, J.~C., \& {Hernquist}, L. 1996, \apj, 464, 641

\bibitem[{{Naab} {et~al.}(2007){Naab}, {Johansson}, {Ostriker}, \&
  {Efstathiou}}]{2007ApJ...658..710N}
{Naab}, T., {Johansson}, P.~H., {Ostriker}, J.~P., \& {Efstathiou}, G. 2007,
  \apj, 658, 710

\bibitem[Oppenheimer et al.(2010)]{2010MNRAS.406.2325O} Oppenheimer, B.~D., 
Dav{\'e}, R., Kere{\v s}, D., Fardal, M., Katz, N., Kollmeier, J.~A., 
\& Weinberg, D.~H.\ 2010, \mnras, 406, 2325 

\bibitem[{{Peng} {et~al.}(2002){Peng}, {Ho}, {Impey}, \&
  {Rix}}]{2002AJ....124..266P}
{Peng}, C.~Y., {Ho}, L.~C., {Impey}, C.~D., \& {Rix}, H.-W. 2002, \aj, 124, 266

 \bibitem[Peng et al.(2010)]{2010arXiv1003.4747P} Peng, Y.-j., et al.\ 2010, 
\apj, 721, 193 
  
  \bibitem[Pizagno et al.(2007)]{2007AJ....134..945P} Pizagno, J., et al.\ 
2007, \aj, 134, 945 

\bibitem[{{Quilis} {et~al.}(2000){Quilis}, {Moore}, \&
  {Bower}}]{2000Sci...288.1617Q}
{Quilis}, V., {Moore}, B., \& {Bower}, R. 2000, Science, 288, 1617

\bibitem[Scannapieco et al.(2009)]{2009MNRAS.396..696S} Scannapieco, C., 
White, S.~D.~M., Springel, V., \& Tissera, P.~B.\ 2009, \mnras, 396, 696 

\bibitem[{{Scarlata} {et~al.}(2007){Scarlata}, {Carollo}, {Lilly}, {Feldmann},
  {Kampczyk}, {Renzini}, {Cimatti}, {Halliday}, {Daddi}, {Sargent},
  {Koekemoer}, {Scoville}, {Kneib}, {Leauthaud}, {Massey}, {Rhodes}, {Tasca},
  {Capak}, {McCracken}, {Mobasher}, {Taniguchi}, {Thompson}, {Ajiki}, {Aussel},
  {Murayama}, {Sanders}, {Sasaki}, {Shioya}, \&
  {Takahashi}}]{2007ApJS..172..494S}
{Scarlata}, C., et al. 2007, \apjs, 172, 494

\bibitem[{{Simha} {et~al.}(2009){Simha}, {Weinberg}, {Dav{\'e}}, {Gnedin},
  {Katz}, \& {Kere{\v s}}}]{2009MNRAS.399..650S}
{Simha}, V., {Weinberg}, D.~H., {Dav{\'e}}, R., {Gnedin}, O.~Y., {Katz}, N., \&
  {Kere{\v s}}, D. 2009, \mnras, 399, 650

\bibitem[{{Sommer-Larsen} {et~al.}(2003){Sommer-Larsen}, {G{\"o}tz}, \&
  {Portinari}}]{2003ApJ...596...47S}
{Sommer-Larsen}, J., {G{\"o}tz}, M., \& {Portinari}, L. 2003, \apj, 596, 47

\bibitem[{{Spergel} {et~al.}(2007){Spergel}, {Bean}, {Dor{\'e}}, {Nolta},
  {Bennett}, {Dunkley}, {Hinshaw}, {Jarosik}, {Komatsu}, {Page}, {Peiris},
  {Verde}, {Halpern}, {Hill}, {Kogut}, {Limon}, {Meyer}, {Odegard}, {Tucker},
  {Weiland}, {Wollack}, \& {Wright}}]{2007ApJS..170..377S}
{Spergel}, D.~N., et al. 2007, \apjs, 170, 377

\bibitem[{{Springel} {et~al.}(2005){Springel}, {Di Matteo}, \&
  {Hernquist}}]{2005MNRAS.361..776S}
{Springel}, V., {Di Matteo}, T., \& {Hernquist}, L. 2005, \mnras, 361, 776

\bibitem[{{Stinson} {et~al.}(2006){Stinson}, {Seth}, {Katz}, {Wadsley},
  {Governato}, \& {Quinn}}]{2006MNRAS.373.1074S}
{Stinson}, G., {Seth}, A., {Katz}, N., {Wadsley}, J., {Governato}, F., \&
  {Quinn}, T. 2006, \mnras, 373, 1074

\bibitem[Teyssier et al.(2011)]{2010arXiv1003.4744T} Teyssier, R., Moore, 
B., Martizzi, D., Dubois, Y., \& Mayer, L.\ 2011, \mnras, 618 

\bibitem[{{Thomas} {et~al.}(2005){Thomas}, {Maraston}, {Bender}, \& {Mendes de
  Oliveira}}]{2005ApJ...621..673T}
{Thomas}, D., {Maraston}, C., {Bender}, R., \& {Mendes de Oliveira}, C. 2005,
  \apj, 621, 673

\bibitem[{{Treu} {et~al.}(2001){Treu}, {Stiavelli}, {M{\o}ller}, {Casertano},
  \& {Bertin}}]{2001MNRAS.326..221T}
{Treu}, T., {Stiavelli}, M., {M{\o}ller}, P., {Casertano}, S., \& {Bertin}, G.
  2001, \mnras, 326, 221

\bibitem[{{Tully} \& {Fisher}(1977)}]{1977A&A....54..661T}
{Tully}, R.~B., \& {Fisher}, J.~R. 1977, \aap, 54, 661

\bibitem[{{van Dokkum} {et~al.}(2009){van Dokkum}, {Kriek}, \&
  {Franx}}]{2009Natur.460..717V}
{van Dokkum}, P.~G., {Kriek}, M., \& {Franx}, M. 2009, \nat, 460, 717

\bibitem[{{Verheijen}(2001)}]{2001ApJ...563..694V}
{Verheijen}, M.~A.~W. 2001, \apj, 563, 694

\bibitem[{{Verheijen} \& {Sancisi}(2001)}]{2001A&A...370..765V}
{Verheijen}, M.~A.~W., \& {Sancisi}, R. 2001, \aap, 370, 765

\bibitem[{{Wadsley} {et~al.}(2004){Wadsley}, {Stadel}, \&
  {Quinn}}]{2004NewA....9..137W}
{Wadsley}, J.~W., {Stadel}, J., \& {Quinn}, T. 2004, New Astronomy, 9, 137

\end{thebibliography}

\clearpage

\end{document}